\documentclass[a4paper,12pt]{article}
\usepackage{axodraw4j}
\usepackage{color}
\usepackage{pstricks}
\usepackage{amssymb,amsmath,epsfig,graphicx,psfrag}
\usepackage{hyperref}

\numberwithin{equation}{section}

\newcommand{\per}{\,}

\newcommand{\pq}[2]{p_{#1}\hspace{-0.ex} q_{#2}}
\newcommand{\qq}[2]{q_{#1}\hspace{-0.ex} q_{#2}}
\newcommand\eqcs{\raisebox{-2mm}{\rlap{\,{\rm cs}\,}} \raisebox{.0ex}{$\,=\,$}}

\newcommand{\as}{\alpha_{\mathrm{S}}}
\newcommand{\bra}[1]{\langle #1 |}

\newcommand{\bs}{\boldsymbol}
\newcommand{\cols}{R}                      
\newcommand{\e}{\varepsilon}
\newcommand{\Ga}{\boldsymbol{\Gamma}}             

\newcommand{\hq}{{\bs q}}                   
\newcommand{\I}{I}
\newcommand{\II}{I\!I}
\newcommand{\III}{I\!I\!I}
\newcommand{\ignore}[1]{}
\newcommand{\J}{\boldsymbol{J}}             
\newcommand{\ket}[1]{| #1 \rangle}

\newcommand{\llbar}{\ell\bar{\ell}}         
\newcommand{\LLbar}{L\bar{L}}               
\newcommand{\lra}{\leftrightarrow}
\newcommand{\M}{{\cal M}}                  
\newcommand{\MIX}{\mathrm{MIX}}            
\newcommand{\ord}[1]{\mathcal{O}\left(#1\right)}

\newcommand{\QED}{\mathrm{QED}}
\newcommand{\qqbar}{{\mathrm{q}\bar{\mathrm{q}}}}
\newcommand{\QQbar}{{\mathrm{Q}\bar{\mathrm{Q}}}}
\newcommand{\Quark}{\mathrm{Q}}           
\newcommand{\quark}{\mathrm{q}}           
\newcommand{\rS}{\widetilde{\mathcal{S}}}           
\newcommand{\cS}{{\cal S}}               

\newcommand{\sla}[1]{#1\hspace{-0.45em}/\hspace{0.em}}
\newcommand{\sy}[2]{\bigl(#1\, #2\bigr)_{sym}}
\newcommand{\T}{\boldsymbol{T}}                   
\newcommand{\tr}{\mathrm{tr}}             
\newcommand{\tri}[1]{{\cal T}^{(d)}_{#1}}   
\newcommand{\ui}{\mathrm{i}}             
\newcommand{\x}{\mathrm{x}}                
\newcommand{\y}{\mathrm{y}}                
\newcommand{\z}{\mathrm{z}}                

\definecolor{dccolor}{rgb}{1 ,0.1,0.7}

\newcommand\dcout{\marginpar{\color{dccolor}$\clubsuit$}\bgroup\markoverwith{\color{dccolor}{\rule[0.4ex]{2pt}{0.8pt}}}\ULon}

\def\lra{\leftrightarrow}

\newcommand\g{g_{\mathrm{S}}}

\newcommand\gq{g}

\def\ep{\epsilon}

\def\beq{\begin{equation}}
\def\eeq{\end{equation}}
\def\beeq{\begin{eqnarray}}
\def\eeeq{\end{eqnarray}}
\def\cm{{\cal M}}
\def\bom#1{{\mbox{\boldmath $#1$}}}
\def\to{\rightarrow}

\newcommand{\la}{\langle}
\newcommand{\ra}{\rangle}

\def\nn{\nonumber}

\def\ID{1 \kern -.45 em 1}

\def\ket#1{|{#1}\ra}
\def\bra#1{\la{#1}|}

\def\ubar{{\overline u}}

\def\cbet0{b_0}

\def\bj{{\bom J}}

\def\btq{{\bom t}}

\def\w0abc{w_{\{ABC\}}}

\begin{document}
\begin{titlepage}
\renewcommand{\thefootnote}{\fnsymbol{footnote}}
\begin{flushright}
FTUV-26-0216.3216
\end{flushright}
\par \vspace{10mm}
\begin{center}
{\Large \bf Four fermion soft emission\\ in QCD hard scattering
}
\end{center}

\par \vspace{2mm}
\begin{center}
{\bf Leandro Cieri}~$^{(a)}$ and {\bf Dimitri Colferai}~$^{(b)}$

\vspace{5mm}

${}^{(a)}$Instituto de F\'{\i}sica Corpuscular, Universitat de Val\`{e}ncia -- Consejo Superior de Investigaciones Cient\'{\i}ficas, Parc Cient\'{\i}fic, E-46980 Paterna, Valencia, Spain

\vspace*{2mm}

${}^{(b)}$
Dipartimento di Fisica e Astronomia,
Universit\`a
di Firenze and\\
INFN Sezione di Firenze,
I-50019 Sesto Fiorentino, 
Florence, Italy \\

\vspace{5mm}

\end{center}

\par \vspace{2mm}
\begin{center} {\large \bf Abstract} \end{center}
\begin{quote}
\pretolerance 10000

We consider the radiation of two distinguishable soft quark-antiquark pairs ($\qqbar\QQbar$) in a generic process for multiparton hard
scattering in QCD. We evaluate the corresponding soft current at tree level in terms of an independent-emission contribution and an irreducible correlation component, which includes strictly non-abelian terms and also terms with an abelian character. The squared current for soft $\qqbar\QQbar$ emission produces colour dipole and colour tripole interactions between the hard-scattering partons, with structures similar to soft gluon-quark-antiquark emission. The colour tripole interactions are odd under charge conjugation and lead to charge asymmetry effects.
We extend our analysis by including QED interactions, the emission of two distinguishable soft lepton-antilepton pairs, and mixed quark-antiquark-lepton-antilepton soft emission.

\end{quote}

\vspace*{\fill}
\begin{flushleft}
$^{(a)}$ email: \href{mailto:lcieri@ific.uv.es}{lcieri@ific.uv.es}\\
$^{(b)}$ email: \href{mailto:dimitri.colferai@unifi.it}{dimitri.colferai@unifi.it}
\end{flushleft}
\end{titlepage}

\vskip 1cm


\setcounter{footnote}{2}

\section{Introduction\label{s:intro}}

The large amount of high-precision data taken at (and much more to come from) the Large Hadron Collider (LHC) demands theoretical predictions at a corresponding high precision.
In the context of quantum chromodynamics (QCD), the theoretical accuracy is typically increased by performing calculations of radiative corrections at higher orders.
The present high-precision frontier is represented by computations at the next-to-next-to-next-to-leading order (N$^3$LO) in the QCD coupling $\as$.

In theories with massless particles, like QCD, higher order scattering amplitudes are typically infrared (IR) divergent, and finite (physical) results are obtained by combining together real and virtual contributions, integrated in soft and collinear regions of phase space. The basic property that allows the cancellation of such IR singularities is their process-independent structure. They are controlled by universal factorization formulae and corresponding soft/collinear factors, which are relevant for both fixed-order and resummed QCD calculations.

The technical effort required to achieve such cancellations increases significantly with the perturbative order. Soft/collinear factorization formulae can be used to organize and greatly simplify the cancellation mechanism of the IR divergences in fixed-order calculations.
The existence of such IR singularities is closely related to the presence of large logarithmic corrections in perturbative scattering amplitudes when evaluated in kinematical regions close to the exclusive boundary of the phase space.
These large logarithms arise from the kinematical unbalance between loop and real radiative corrections in the soft and collinear regions of the phase space. Soft/collinear factorization formulae and the corresponding singular factors are the basic ingredients for the explicit computation and resummation of these large logarithmic contributions.


The singular factors at ${\cal O}(\as)$ and ${\cal O}(\as^2)$  for soft and collinear factorization of scattering amplitudes have been known for a long time. The explicit knowledge of soft/collinear factorization at ${\cal O}(\as)$ has been essential to devise fully general (process-independent and observable-independent) methods to carry out next-to-leading order (NLO) QCD calculations (see, e.g., Refs.~\cite{Frixione:1995ms, csdip}). Similarly, the knowledge of soft/collinear factorization formulae at ${\cal O}(\as^2)$ \cite{Campbell:1997hg, Catani:1998nv, Bern:1998sc, Kosower:1999rx, Bern:1999ry,Catani:1999ss, Catani:2000pi, Czakon:2011ve, Bierenbaum:2011gg, Catani:2011st, Sborlini:2013jba} is exploited to develop methods (see, e.g., the review in Ref.~\cite{Heinrich:2020ybq}) at the next-to-next-to-leading order (NNLO). Soft/collinear factorization up to ${\cal O}(\as^2)$ contributes to resummed calculations up to next-to-next-to-leading logarithmic (NNLL) accuracy (see, e.g., Refs.~\cite{Becher:2014oda, Luisoni:2015xha}).

Soft and collinear factorization at ${\cal O}(\as^3)$ can be used in the context of N$^3$LO calculations and of resummed calculations at N$^3$LL accuracy. The process-independent singular factors for the various collinear limits at  ${\cal O}(\as^3)$ are presented in Refs.~\cite{Catani:2011st,DelDuca:1999iql, Birthwright:2005ak, DelDuca:2019ggv, Catani:2003vu, Sborlini:2014mpa, Badger:2015cxa, Bern:2004cz, Badger:2004uk, Duhr:2014nda}. Soft factorization of scattering amplitudes at ${\cal O}(\as^3)$ requires the study of various tree-level and loop contributions. Triple soft-gluon radiation at tree level is studied in Ref.~\cite{Catani:2019nqv}. Double soft emission at one loop level has been considered in Refs.~\cite{Zhu:2020ftr,Catani:2021kcy}. Single soft-gluon radiation at two loop order is examined in detail in Refs.~\cite{Badger:2004uk, Li:2013lsa, Duhr:2013msa, Dixon:2019lnw}. Recently soft gluon-quark-antiquark radiation at tree level has been studied in Refs.~\cite{DelDuca:2022noh,Catani:2022hkb}.

Very recently the tree-level soft emission for two pairs of quarks was presented in Ref.~\cite{Chen:2024hvp}.
Here we reconsider this process, by performing a completely independent calculation in QCD (and then extending the results to QED interactions).
More precisely, we consider the singular behaviour of tree-level QCD scattering amplitudes in the limit in which two distinguishable quark-antiquark pairs are soft.
The singularity is controlled in factorized form by a current for soft four-parton radiation from hard partons. 
We discuss some properties of quadruple soft radiation at the level of both scattering amplitudes and squared amplitudes. In agreement with Ref.~\cite{Chen:2024hvp}, we find that radiation of two quark-antiquark pairs produces both colour dipole and non-abelian colour tripole correlations between the hard partons in generic processes with three or more hard partons. However, the corresponding kinematical coefficients do not agree numerically, even when applied to colour-singlet hard amplitudes.

We analyse in some detail the charge asymmetry effects induced by tripole correlations.
We also specialize our results to processes with two and three hard partons, where matrix factorization reduces to a simpler c-number factorization. Finally, our results can be adapted to mixed QED$\times$QCD interactions where photons can be exchanged between quark lines, and also soft lepton-antilepton pairs are radiated in the final state.

The outline of the paper is as follows.
In Sect.~\ref{s:sfsc} we first introduce our notation and recall the soft factorization formula for scattering amplitudes. Subsequently, we present the calculation of the tree-level current for emission of two soft quark-antiquark pairs in a generic hard-scattering process. The result for the current has an irreducible correlation component that includes contributions
with both abelian and non-abelian characters.
In Sect.~\ref{s:tlsc} we consider soft factorization of
squared amplitudes and we compute the squared current for radiation of two soft
quark-antiquark pairs. The squared
current leads to irreducible colour dipole and colour tripole interactions. The colour tripole
interactions are odd under charge conjugation and produce charge asymmetry eﬀects
between the soft quark and antiquark.
In Sect.~\ref{s:e23h} we consider the specific applications to
processes with two and three hard partons and, in particular, we discuss the structure of the
corresponding charge asymmetry contributions.
In Sect.~\ref{s:qed} we generalize our QCD results to the cases of QED and mixed
QCD$\times$QED radiative corrections for a generic emission of two soft fermion-antifermion pairs, where the fermions are either quarks or leptons.
A brief summary of our results is presented in Sect.~\ref{s:sum}.

\section{Soft factorization and soft currents\label{s:sfsc}}

In this section we first introduce our notation, mostly
following the notation that is also used in Refs.~\cite{Catani:2019nqv, Catani:2021kcy}
(more details can be found therein). 
We also briefly recall 
the factorization properties of scattering amplitudes in the soft limit and the known 
tree-level results for the emission of a soft quark-antiquark pair.
Then we present and discuss our results of the soft current for the emission of two distinguishable quark-antiquark pairs at tree level.

\subsection{Soft factorization of scattering amplitudes\label{s:sfsa}}

We study the soft behaviour of a generic scattering amplitude $\M$ whose
external-leg particles are on-shell and with physical spin polarizations. In our notation
all external particles of $\M$ are treated as `outgoing' particles
(although they can be initial-state and final-state physical particles), with corresponding outgoing momenta and quantum numbers (e.g., colour, spin and flavour).
The perturbative evaluation of $\M$ is performed by using dimensional regularization in
$d=4-2\epsilon$ space-time dimensions, and $\mu$ is the dimensional regularization scale. Specifically, we use conventional dimensional regularization (CDR), with $d-2$ spin polarization states for on-shell gluons (and photons) 
and 2 polarization states for on-shell massless quarks or antiquarks (and massless leptons). 

We consider the behaviour of 
$\M$ in the
kinematical configuration where one or more of the momenta of the external-leg
massless particles become soft. 
We denote the
soft momenta by $q_\ell^\mu$ ($\ell=1,\dots,N$, with $N$, the total
number of soft particles), while the momenta of the hard particles in $\M$ are
denoted by $p_i^\mu$ (in general they are not massless and $p_i^2 \equiv m_i^2 \neq 0$)
In this kinematical configuration, $\M(\{q_\ell\},
\{p_i\})$ becomes singular. The dominant singular behaviour is given by the
following factorization formula in colour space 
\cite{Catani:1999ss, Bern:1999ry, Catani:2000pi}:
\begin{equation}\label{1gfact}
  \ket{\M(\{q_\ell\}, \{p_i\})} \simeq
\J(q_1,\cdots,q_N) \; \ket{\M (\{p_i\})} \;,
\end{equation}
Here $\M (\{p_i\})$ is
the scattering amplitude that is obtained from the original amplitude
$\M(\{q_\ell\}, \{p_i\})$ by simply removing the soft external legs.  The factor
$\J$ is the soft current for multi-particle radiation from the scattering
amplitude.

At the formal level the soft behaviour of $\M(\{q_\ell\}, \{p_i\})$ is specified
by performing an overall rescaling of all soft momenta as $q_\ell \to \xi
q_l$ (the rescaling parameter $\xi$ is the same for each soft momentum $q_\ell$)
and by considering the limit $\xi\to 0$.  In this limit, the amplitude is
singular and it behaves as $(1/\xi)^N$ (modulo powers of $\ln \xi$ from loop
corrections). This dominant singular behaviour is embodied in the soft current
$\J$ on the right-hand side of Eq.~(\ref{1gfact}).  In this equation the symbol
$\simeq$ means that on the right-hand side we neglect contributions that
are less singular than $(1/\xi)^N$ in the limit $\xi\to 0$.

The soft current $\J(q_1,\cdots,q_N)$ in Eq.~(\ref{1gfact}) depends on the
momenta, colours and spins of both the soft and hard partons in the scattering
amplitude (although, the hard-parton dependence is not explicitly denoted in the
argument of $\J$). However this dependence entirely follows from the
external-leg content of $\M$, and the soft current is completely independent of
the internal structure of the scattering amplitude. In particular, we remark
that the factorization in Eq.~(\ref{1gfact}) is valid \cite{Bern:1999ry,
  Catani:2000pi, Feige:2014wja} at arbitrary perturbative orders in the loop expansion of the
scattering amplitude. 
Therefore on both sides of Eq.~(\ref{1gfact}) the scattering amplitudes have the loop
expansion ${\ket \M}= {\ket {\M^{(0)}}}+{\ket {\M^{(1)}}}+\dots$, where $\M^{(0)}$
is the contribution to $\M$ at the lowest perturbative order,
$\M^{(1)}$ is the one-loop contribution, and so forth.
Correspondingly, we have $\J= \J^{(0)}+\J^{(1)}+\dots$,
where $\J^{(n)}$ is the contribution to $\J$ at the $n$-th loop accuracy.  In
 the following sections of this paper we limit ourselves to considering 
explicit expressions of 
only tree-level currents $\J^{(0)}$ and, for the sake of
simplicity, we simply denote them by $\J$ (removing the explicit superscript
$(0)$).

Considering the emission of soft QCD partons,
the all-loop 
current $\J$ in Eq.~(\ref{1gfact}) is an operator that acts
from the colour+spin space of $\M(\{p_i\})$ to the enlarged space of
$\M(\{q_\ell\}, \{p_i\})$.  In particular, soft radiation produces colour
correlations.  To take into account the colour structure we use the colour (+
spin) space formalism of Ref.~\cite{csdip}. The scattering amplitude
$\M_{s_1 s_2 \dots}^{c_1 c_2\dots}$
depends on the colour ($c_i$) and
spin 
($s_i$)
indices of its external-leg partons. This dependence is
embodied in a vector $\ket{\M}$ in colour+spin space through the definition
(notation)
\begin{equation}\label{Mstate}
  \M_{s_1 s_2 \dots}^{c_1 c_2 \dots} \equiv
  \big(\bra{c_1,c_2,\cdots}\otimes\bra{s_1,s_2,\cdots}\big) \;
  \ket{\M} \;\;,
\end{equation}
where 
$\{ \,\ket{c_1,c_2,\cdots}\otimes\ket{s_1,s_2,\cdots} \} = \{ \,
\ket{c_1,s_1;c_2,s_2,\cdots} \}$ 
is an orthonormal basis of abstract
vectors in colour+spin space.

In colour space the colour correlations produced by soft-gluon emission are
represented by associating a colour charge operator $\T_i$ to the emission of a
gluon from each parton $i$. If the emitted gluon has colour index $a$
($a=1,\dots,N_c^2-1$, for $SU(N_c)$ QCD with $N_c$ colours) in the adjoint
representation, the colour charge operator is $\T_i \equiv \bra{a} \,T_i^a$ and
its action onto the colour space is defined by
\begin{equation}\label{defT}
  \bra{a,c_1,\cdots,c_i,\cdots,c_m}\,\T_i\,\ket{b_1,\cdots,b_i,\cdots,b_m} \equiv
  \delta_{c_1 b_1} \cdots (T^a)_{c_i b_i} \cdots \delta_{c_m b_m} \;,
\end{equation}
where the explicit form of the colour matrices $T^a_{c_i b_i}$ depends on the
colour representation of the parton $i$:
\begin{align}
 (T^a)_{b c} &= \ui f^{bac} && \text{(adjoint representation) 
if $i$ is a gluon,} &&\nonumber \\
  (T^a)_{\alpha\beta} &=  t^a_{\alpha\beta} &&
  \text{(fundamental representation with $\alpha,\beta=1,\dots,N_c$)
    if $i$ is a quark,}  &&\nonumber \\
  (T^a)_{\alpha\beta} &= -t^a_{\beta\alpha}   &&
  \text{if $i$ is an antiquark.} &&\nonumber &&\label{Tcs}
\end{align}
We normalize the colour matrices such that $[T_i^a , T_j^b] = i f^{abc} T_i^c \delta_{ij}$
and ${\rm Tr} (t^a t^b) = T_R \,\delta_{ab}$ with $T_R=1/2$.
We also use the notation
$T_i^a T_k^a \equiv \T_i \cdot \T_k$ and $\T_i^2 = C_i$, where $C_i$ is the
quadratic Casimir coefficient of the colour representation, with the
normalization $C_i=C_A=N_c$ if $i$ is a gluon and $C_i=C_F=(N_c^2-1)/(2N_c)$ if
$i$ is a quark or antiquark.

Note that each `amplitude vector' $\ket{\M}$ is an overall colour-singlet state.
Therefore, colour conservation is simply expressed by the relation
\begin{equation}
\label{colcons}
\sum_i \;\T_i \;\ket{\M} = 0 \;\;,
\end{equation}
where the sum extends over all the external-leg partons $i$ of the amplitude $\M$. 
For subsequent use, we also introduce the shorthand notation
\begin{equation}
\label{csnotation}
\sum_i \;\T_i \;\eqcs \;0 \;\;,
\end{equation}
where the subscript CS in the symbol $\eqcs$ means that the equality between the
terms in the left-hand and right-hand sides of the equation is valid if these
(colour operator) terms act (either on the left or on the right) onto
colour-singlet states.

\subsection{Tree-level currents\label{s:tlc}}

The tree-level current $\bj(q)$ for the emission of a soft quark-antiquark ($\qqbar$) pair by tree-level QCD interactions was studied
in Ref.~\cite{Catani:1999ss}. Using our notation, the QCD current for radiation of a soft quark and antiquark at tree level is \cite{Catani:2021kcy}
\begin{equation}\label{Jqa}
\bj(q_1,q_2) 
=
- \left( \g\,\mu^\ep\right)^2 \,
\sum_i 
\,\btq^c \;T^c_i
\; \xi_i(1,2)\;\;,
\end{equation}
in terms of the kinematical coefficient $\xi_i(1,2)$ and the fermionic current $j^\nu(1,2)$,
\begin{equation}\label{fercur}
\xi_i(1,2) = \frac{p_i \cdot j(1,2)}{p_i \cdot q_{12}}
\;, \qquad
j^\nu(1,2) \equiv \frac{\ubar(q_1)\, \gamma^\nu \,v(q_2)}{q_{12}^2} \;\;,
\quad\;\;\; \quad q_{12} \equiv q_1+q_2 \;\;.
\end{equation}
The soft quark and antiquark have momenta $q_1^\nu$ and $q_2^\nu$, respectively,
and $u(q)$ and $v(q)$ are the customary Dirac spinors. The spin indices ($s_1$ and
$s_2$) and the colour indices ($\alpha_1$ and $\alpha_2$) of the 
quark and antiquark are embodied in the colour+spin space notation of 
Eq.~(\ref{Jqa}). 
Considering the projection 
$(\bra{\alpha_1,\alpha_2} \otimes \bra{s_1,s_2} \,) \,\bj(q_1,q_2)
\equiv J^{\alpha_1,\alpha_2}_{s_1,s_2}(q_1,q_2)$
of the current onto its colour and spin indices, we have
$(\bra{\alpha_1,\alpha_2} \otimes \bra{s_1,s_2} \,) \,\btq^c \;
\ubar(q_1)\, \gamma^\nu \,v(q_2) = t^c_{\alpha_1\alpha_2} \,
\ubar_{(s_1)}(q_1)\, \gamma^\nu \,v_{(s_2)}(q_2)$.

\subsubsection{The tree-level current for soft $\qqbar\QQbar$ emission\label{s:qqQQc}}

\begin{figure}
    \centering
    \includegraphics[width=0.95\linewidth]{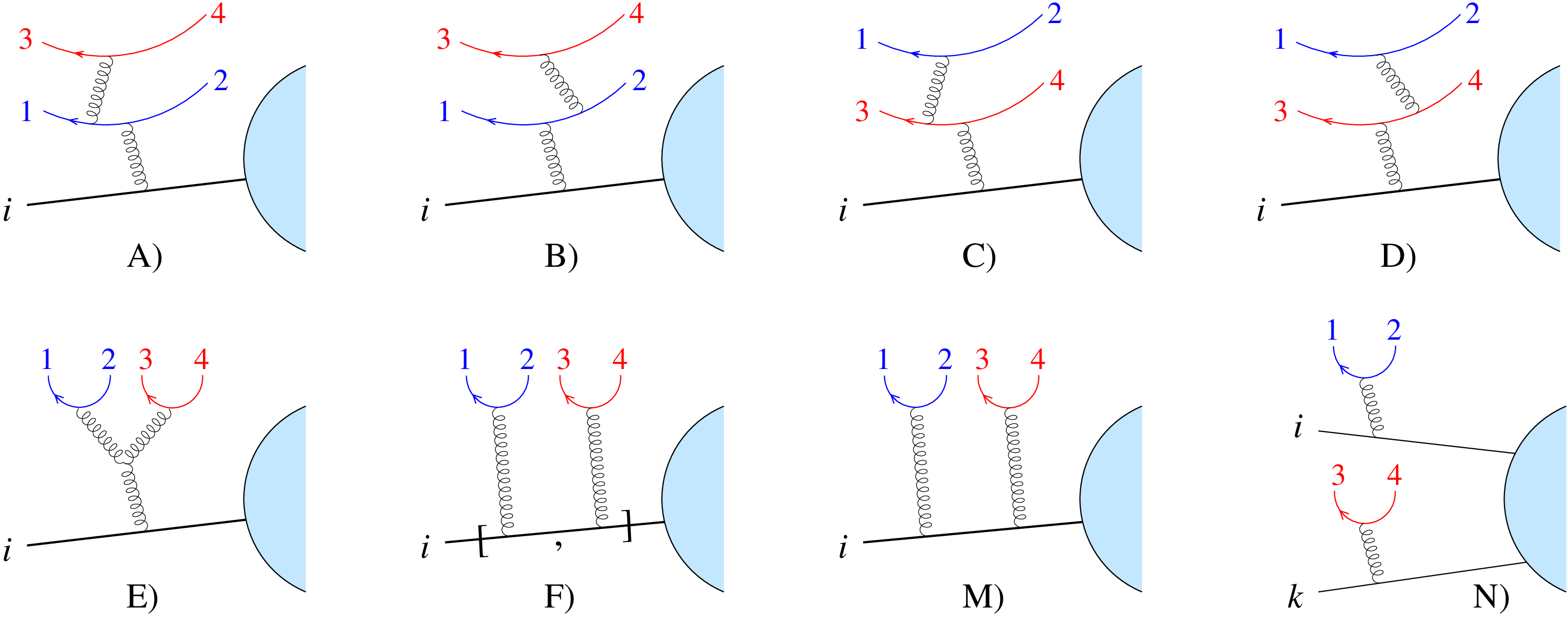}
    \caption{QCD diagrams contributing to the $\qqbar\QQbar$ soft emission current.
    The $\qqbar$ pair is shown in blue and the $\QQbar$ pair in red. Diagrams from A to E correspond to soft quark emission stemming from a single gluon leg attached to a hard parton. In diagrams M and N each soft pair stems from a different gluon attached to a hard leg. The sum of diagrams of type M and N yields the symmetric product of the quark-antiquark currents $\sy{\bj_{12}}{\bj_{34}}$ of Eq.~\eqref{J1234} plus the commutator of the same currents attached on the same leg, represented by diagram F. Therefore, the irreducible correlation $\Ga$ is given by the sum of diagrams A-F.}
    \label{f:qqQQ}
\end{figure}

The relevant diagrams for the emission of two soft {\em distinguishable} quark-antiquark pairs are depicted in fig.~\ref{f:qqQQ}. We denote $\qqbar$ the first pair with particle labels 1, 2 while $\QQbar$ denotes the second pair with particle labels 3, 4. We perform the calculation in light-cone gauge, introducing an auxiliary light-like vector $n$ in terms of which the gluon propagator is given by
\begin{equation}
  D_{\mu\nu}(k) = \frac{\ui}{k^2} d_{\mu\nu}(k) \;, \qquad
  d_{\mu\nu}(k) = \sum_{\lambda}\epsilon_\mu^{(\lambda)}(k)^*\epsilon_\nu^{(\lambda)}(k)
  = -g_{\mu\nu}+\frac{k^\mu n^\nu+n^\mu k^\nu}{k\cdot n} \;.
\end{equation}

Each pair stems from a soft gluon. Following Ref.~\cite{Catani:1998nv}, the contributions from diagrams M and N can be decomposed as the symmetric product of the two single-pair currents --- $\bj_\qqbar\equiv\bj(q_1,q_2)$ and $\bj_\QQbar\equiv\bj(q_3,q_4)$ --- plus a commutator of single-pair currents on the same hard leg, represented by diagram F. In practice, the total soft current can be presented in the form
\begin{equation}\label{J1234}
  \bj(q_1,q_2,q_3,q_4) =
  \sy{\bj(q_1,q_2)}{\bj(q_3,q_4)} + \Ga(q_1,q_2,q_3,q_4)
\end{equation}
and we have introduced the symbol $(\dots)_{sym}$ to denote symmetrized products.
The symmetrized product of two colour space operators $A$ and $B$ is defined as
\begin{equation}\label{symprod}
  \sy{A}{B} = \frac12(AB+BA) \;.
\end{equation}
The right-hand side of Eq.~(\ref{J1234}) has the structure of an expansion in irreducible
correlations, which is analogous to the structure of the two-gluon, three-gluon and gluon-quark-antiquark
soft currents in Refs.~\cite{Catani:1999ss}, \cite{Catani:2019nqv} and \cite{Catani:2022hkb} respectively.
The first term in the right-hand side of Eq.~(\ref{J1234}) represents the `independent'
(though colour-correlated) emission of the two soft quark-antiquark pairs
from the hard partons. The term $\Ga(q_1,q_2,q_3,q_4)$ is strictly an irreducible correlation contribution to soft $\qqbar\QQbar$ emission, and can be decomposed into three independent colour structures:
\begin{align}\label{G1234}
  &\Ga(q_1,q_2,q_3,q_4) \eqcs \left(\g\,\mu^\ep\right)^4 \,
  \sum_i T^c_i \,\sum_{\sigma\in\{\x,\y,\z\}} \cols_{(\sigma)}^c \gamma_i^{(\sigma)}(q_1,q_2,q_3,q_4) \\
  &\cols_{(\x)}^c = \frac12 \{t^a,t^c\}_{\alpha_1\alpha_2} t^a_{\alpha_3 \alpha_4} \;,\quad
  \cols_{(\y)}^c = \frac12 t^a_{\alpha_1 \alpha_2} \{t^a,t^c\}_{\alpha_3\alpha_4} \;,\quad
  \cols_{(\z)}^c = \frac{\ui}{2} f^{abc} t^a_{\alpha_1 \alpha_2} t^b_{\alpha_3\alpha_4} \label{R123} \\
  &\gamma_i^{(\x)} = \frac{j_\nu(3,4)}{q_{1234}^2\;p_i\cdot q_{1234}}\;
  \ubar_1\left(\frac{\gamma^\nu\sla{q}_{134}\sla{p}_i}{q_{134}^2}
  -\frac{\sla{p}_i\sla{q}_{234}\gamma^\nu}{q_{234}^2}\right)v_2 \;, \qquad
  \gamma_i^{(\y)} = \gamma_i^{(\x)}|_{\substack{1\lra 3\\2\lra 4}} \label{g12} \\
  &\gamma_i^{(\z)} = \frac{j_\nu(3,4)}{q_{1234}^2\;p_i\cdot q_{1234}}\;\bigg\{
  \ubar_1\left(\frac{\gamma^\nu\sla{q}_{134}\sla{p}_i}{q_{134}^2}
  +\frac{\sla{p}_i\sla{q}_{234}\gamma^\nu}{q_{234}^2}\right)v_2 \nonumber \\
  &\qquad\qquad +j_\mu(1,2)\left[4p_i^\mu q_{12}^\nu + 2p_i\cdot q_{34} \, g^{\mu\nu}
  + \frac{q_{1234}^2}{p_i\cdot q_{12}} p_i^\mu p_i^\nu \right]\bigg\}
  - \left( \substack{1\lra 3\\2\lra 4} \right) \;. \label{g3} 
\end{align}
The colour structures `$\x$' and `$\y$' can be considered of abelian type (they survive in the same form also for an abelian theory), while the `$\z$' structure is purely non abelian.

Concerning the kinematical coefficients (where we use the shorthand notation $q_{134}\equiv q_1+q_3+q_4$ and so on), we note that the dependence on the auxiliary gauge vector $n$ disappears when the current $\J$ acts on colour singlet states. Therefore the expressions given above are gauge-independent.

\section{Tree-level squared currents\label{s:tlsc}}

Using the colour+spin space notation of Sect.~\ref{s:sfsa}, 
the squared amplitude $|\M|^2$ (summed
over the colours and spins of its external legs) is written as follows
\beq
\label{squared}
|\M|^2 = \langle{\M} \ket{\,\M} \;\;.
\eeq
Accordingly, the square of the soft factorization formula (\ref{1gfact})
gives
\beq
\label{softsquared}
| \M(\{q_\ell\}, \{p_i\}) |^2 \simeq 
\bra{\M (\{p_i\})} \;| \J(q_1,\cdots,q_N) |^2 \;\ket{\M (\{p_i\})} \;,
\eeq
where
\beq
\label{spin}
| \J(q_1,\cdots,q_N) |^2 = \left[ J^{c_1 \dots c_N}_{s_1 \dots s_N}(q_1,\cdots,q_N) 
\right]^\dagger J^{c_1 \dots c_N}_{s_1 \dots s_N}(q_1,\cdots,q_N) \;\;,
\eeq
and, analogously to Eq.~(\ref{1gfact}), the symbol $\simeq$ means that we neglect contributions that are subdominant in the soft limit.
The squared current $| \J(q_1,\cdots,q_N) |^2 $, which is summed over the colours 
$c_1 \dots c_N$ and spins $s_1 \dots s_N$ of the soft partons, is still a colour operator that depends on the colour charges of the hard partons in $\M (\{p_i\})$. These colour charges produce colour correlations and,
therefore, the right-hand side of Eq.~(\ref{softsquared}) is not proportional to
$|\M(\{p_i\})|^2$ in the case of a generic scattering amplitude\footnote{Colour correlations can be simplified in the case of scattering amplitudes with two and three hard partons (see Sect.~\ref{s:e23h}).}. 

The computation of the squared current in Eq.~(\ref{spin}) involves the sum over the physical spin polarization vectors of the soft gluons. These polarization vectors are gauge dependent, but the action of $| \J |^2 $ onto colour singlet states is fully gauge invariant,
since the gauge dependent contributions cancel as a consequence of the conservation of the soft current (see, e.g., Ref.~\cite{Catani:2019nqv}).

We recall the known results of the squared current for emission of a soft $\qqbar$  pair \cite{Catani:1999ss}:
\begin{equation}\label{Jqasq}
  |\bj(q_1,q_2)|^2 \equiv  \bj(q_1,q_2)^\dagger\; \bj(q_1,q_2)
  \eqcs \left( \g\,\mu^\ep\right)^4 T_R \sum_{i,k} \T_i\cdot \T_k \; {\cal I}_{ik}(q_1,q_2)\;,
\end{equation}
where
\begin{equation}
 {\cal I}_{ik}(q_1,q_2) = \frac{p_i\cdot q_1\; p_k\cdot q_2+p_i\cdot q_2\;
   p_k\cdot q_1-p_i\cdot p_k \;q_1\cdot q_2}%
 {(q_1\cdot q_2)^2\;p_i\cdot q_{12}\; p_k\cdot q_{12}} \label{I12}
\end{equation}
is symmetric in the exchange $(1\lra 2)$ or $(i\lra k)$.

The colour charge dependence of the squared current in Eq.~(\ref{Jqasq}) 
is given in terms of dipole operators $\T_i\cdot \T_k = \sum_a T_i^a  T_k^a$.
The insertion of the dipole operators in the factorization formula (\ref{softsquared})
produces colour correlations between two hard partons ($i$ and $k$) in $\M(\{p_i\})$.

Using colour charge conservation (see Eq.~\eqref{colcons}), the $\qqbar$ squared
current in Eq.~\eqref{Jqasq} 
can be rewritten as follows
\begin{align}
  |\bj(q_1,q_2)|^2 &\eqcs - \left( \g\,\mu^\ep\right)^4 T_R
  \sum_{i<k} \T_i\cdot \T_k \, w_{ik}(q_1,q_2) \;\;,\label{j12q} \\
  w_{ik}(q_1,q_2) &= {\cal I}_{ii}(q_1,q_2)+{\cal I}_{kk}(q_1,q_2)
  -{\cal I}_{ik}(q_1,q_2)-{\cal I}_{ki}(q_1,q_2) \;\;. \label{w12}
\end{align}
As noticed in Ref.~\cite{Catani:2021kcy}, the expression in 
Eq.~(\ref{j12q}) have a more straightforward physical interpretation,
since the kinematical function
$w_{ik}(q_1,q_2)$ in Eq.~(\ref{w12}) is directly related to the intensity of soft radiation
from two distinct hard partons, $i$ and $k$, in a colour singlet configuration
(see Sect.~\ref{s:e2h}).

\subsection{The squared current for soft $\qqbar\QQbar$ radiation}\label{s:qqQQsc}

We compute the square of the tree-level $\qqbar\QQbar$ current by using the structure in Eq.~\eqref{J1234} and we obtain the following three contributions:
($I$) the square of the `uncorrelated' current
$\sy{\bj(q_1,q_2)}{\bj(q_3,q_4)}$, ($\II$) the product of the `uncorrelated' current
and the irreducible current $\Ga(q_1,q_2,q_3,q_4)$, ($\III$) the square of the irreducible current.

The square of the `uncorrelated' current $\sy{\bj(q_1,q_2)}{\bj(q_3,q_4)}$
can be written in terms of the symmetric product
of the squared currents $|\bj(q_1,q_2)|^2$ and $|\bj(q_3,q_4)|^2$, and an additional
irreducible term $W^{(\I)}$, which involves colour dipole correlations. We obtain
\begin{align}\label{Jsq1}
  &|\sy{\bj(q_1,q_2)}{\bj(q_3,q_4)}|^2
  \eqcs \sy{|\bj(q_1,q_2)|^2}{|\bj(q_3,q_4)|^2} + W^{(\I)}(q_1,q_2,q_3,q_4) \;, \\
  &W^{(\I)}(q_1,q_2,q_3,q_4) = -\left( \g\,\mu^\ep\right)^8
  T_R^2 C_A \sum_{i,k} \T_i\cdot \T_k \; \cS^{(\I)}_{ik}(q_1,q_2,q_3,q_4) \;,
   \label{W1}\\
  &\cS^{(\I)}_{ik}(q_1,q_2,q_3,q_4) =
  \frac12 {\cal I}_{ik}(q_1,q_2){\cal I}_{ii}(q_3,q_4)
  +\frac12 {\cal I}_{ii}(q_1,q_2){\cal I}_{ik}(q_3,q_4)
  -\frac34 {\cal I}_{ik}(q_1,q_2){\cal I}_{ik}(q_3,q_4) \;,
\end{align}
which is symmetric in any of the exchanges $(1\lra 2)$, $(3\lra 4)$ or
$\left(\substack{1\lra 3\\2\lra 4}\right)$.

The product of the `uncorrelated' current and the irreducible current leads to colour correlations that involve dipoles and tripoles. We have
\begin{align}
  &\sum\sy{\bj(q_1,q_2)}{\bj(q_3,q_4)}^\dagger\,\Ga(q_1,q_2,q_3,q_4) + \text{h.c.}
  = W^{(\II)}(q_1,q_2,q_3,q_4)
  \nonumber \\
  &\eqcs -\left( \g\,\mu^\ep\right)^8 T_R^2 \left[
    C_A \sum_{i,k}\T_i\cdot \T_k \; \cS^{(\II)}_{ik}(q_1,q_2,q_3,q_4)
    + \sum_{i,k,m} \tri{ikm} \;\cS_{ikm}(q_1,q_2,q_3,q_4)\right] \;, \label{W2}
\end{align}
where `h.c' denotes the hermitian-conjugate contribution, and
we have defined the (hermitian) $d$-type colour tripole  $\tri{ikm}$ as
\begin{equation}\label{tripoli}
  \tri{ikm} \equiv \sum_{a,b,c} d^{abc}\,T_i^a \,T_k^b \,T_m^c \;, \qquad
  d^{abc} = \frac{1}{T_R} \,{\rm Tr}\left( \left\{t^a, t^b \right\} t^c \right) \;.
\end{equation}
The complete symmetry of $d^{abc}$ causes $\tri{ikm}$ to be completely symmetric in the permutations of its indices $i, k, m$.

The kinematical coefficient $\cS_{ikm}$ of the tripole contribution to Eq.~(\ref{W2}) is given by
\begin{align}\label{Sikm}
 &\cS_{ikm}(q_1,q_2,q_3,q_4) = \rS_{ikm}(q_1,q_2,q_3,q_4) + (3\lra 4)
  - (1\lra 2) + \left(\substack{1\lra 3\\2\lra 4}\right) \\
 &\rS_{ikm}(q_1,q_2,q_3,q_4) = 
  \frac{
   2 \, q_3 q_4\,\tr(p_i\, q_2\, p_k\, q_1\, p_m\, q_{134})
  -4 \, p_m q_4\,\tr(p_i\, q_2\, p_k\, q_1\, q_3\, q_{134})
  }{
  q_{1234}^2 \per q_{12}^2 \per (q_{34}^2)^2 \per q_{134}^2
  \per \pq{i}{1234} \per \pq{k}{12} \per \pq{m}{34}} \;. \label{rSikm}
\end{align}
Note that, in order to adopt a more concise notation, we have suppressed the ``slash'' on momenta inside the Dirac traces:
$\tr(p_m\,q_4)\equiv \tr(\sla{p}_m\,\sla{q}_4)$ etc.; we have also suppressed the ``dot'' in the scalar products: $\pq{k}{12}\equiv p_k\cdot q_{12}$ and so on. The expansion of traces into sums of products of scalar products of momenta is straightforward --- it can be implemented automatically --- but yields lengthy expressions, therefore we prefer to keep them unexpanded. Furthermore, we adopt the convention that adding expressions with exchanged momenta applies on the whole quantity to the left:
\begin{align}\nonumber
  &\rS_{ikm}(q_1,q_2,q_3,q_4) + (3\lra 4)
  - (1\lra 2) + \left(\substack{1\lra 3\\2\lra 4}\right)
 \equiv \\
  &\equiv \bigg[ \Big[ \rS_{ikm}(q_1,q_2,q_3,q_4) + (3\lra 4) \Big]
  - (1\lra 2) \bigg] + \left(\substack{1\lra 3\\2\lra 4}\right) \;.
  \label{defExch}
\end{align}
We shall adopt such conventions also in the following expressions.

The function $\cS_{ikm}$ in Eq.~\eqref{Sikm} is obviously symmetric in the ``pair exchange'' $\left(\substack{1\lra 3\\2\lra 4}\right)$ but it is antisymmetric in the simultaneous exchange of quarks with antiquarks, i.e., $\left(\substack{1\lra 2\\3\lra 4}\right)$
(but there is no (anti)symmetry in the exchange of only one quark$\lra$antiquark).
As discussed in Ref.~\cite{Catani:2021kcy}, such
antisymmetric dependence is related to charge asymmetry effects that are distinct features
of the radiation of quarks and antiquarks that depend on the fully-symmetric
colour tensors $d^{abc}$, which is odd under charge conjugation.

The kinematical coefficient $\cS^{(\II)}_{ik}$ of the dipole contribution in Eq.~(\ref{W2}) can be expressed in terms of the function $\rS_{ikm}$ in Eq.~\eqref{rSikm} and of the new function
\begin{align}
  \rS_{ik}^{(\II)}&(q_1,q_2,q_3,q_4)
  =\frac1{
  (q_{12}^2\per q_{34}^2)^2 \per \pq{i}{1234}\per\pq{i}{12}\per\pq{k}{34}
  }
  \nonumber \\
 &\quad\times  \bigg\{
  \frac12\left(\frac1{\pq{i}{34}}-\frac1{\pq{i}{12}}\right)
  \tr(p_i\,q_1\,p_i\,q_2)\,\tr(p_i\,q_3\,p_k\,q_4)
  \nonumber \\
  &\quad + \frac{2}{q_{1234}^2}\Big[
  \tr(p_i\,q_3\,p_k\,q_4)\,\tr(p_i\,q_1\,q_{34}\,q_2)
  -\tr(p_i\,q_1\,p_i\,q_2)\,\tr(p_k\, q_3\,q_{12}\,q_4)
  \nonumber \\
 &\quad + 2\,p_i(q_{12}-q_{34})\big[
    p_k q_3 \,\tr(p_i\,q_1\,q_4\,q_2) + p_k q_4 \,\tr(p_i\,q_1\,q_3\,q_2)
  - q_3 q_4\, \tr(p_i\,q_1\,p_k\,q_2)
  \big] \Big] \bigg\}
\end{align}
in this way:
\begin{align}\label{cSII}
 \cS^{(\II)}_{ik}(q_1,q_2,q_3,q_4) = 
 \left\{\frac12 \left[\rS_{iik}-\rS_{iki}+(1\lra 2)+(3\lra 4)\right]
 +\rS_{ik}^{(\II)} \right\}
 + \left(\substack{1\lra 3\\2\lra 4}\right) \;.
 \end{align}
It is symmetric in any of the exchanges $(1\lra 2)$, $(3\lra 4)$ or
$\left(\substack{1\lra 3\\2\lra 4}\right)$.

The square of the irreducible current gives
\begin{align}
  &\Ga^\dagger(q_1,q_2,q_3,q_4)\,\Ga(q_1,q_2,q_3,q_4) = W^{(\III)}(q_1,q_2,q_3,q_4) \nonumber \\
  &\eqcs -\left( \g\,\mu^\ep\right)^8 T_R^2 \sum_{i,k} \T_i\cdot \T_k
  \left\{
  K_\x\;\cS^{(\III,\x)}_{ik} + K_\y\;\cS^{(\III,\y)}_{ik} + K_\z\;\cS^{(\III,\z)}_{ik}
  \right\} \;, \label{W3}
\end{align}
where the colour coefficients $K_\sigma$ read
\begin{equation}
    K_\x = C_F - \frac{C_A}{4} \;, \quad
    K_\y = 4 C_F - \frac{3}{2}C_A \;, \quad
    K_\z = \frac{C_A}{4}\;,
\end{equation}
and the kinematical function $\cS^{(\III,\sigma)}_{ik}$ are given by
\begin{align}
  &\cS^{(\III,\x)}_{ik} = \left.
  \rS^{(\x)}_{ik}\right|_{\sigma=-1} + (1\lra 2) + (3\lra 4)
  + \left(\substack{1\lra 3\\2\lra 4}\right) \label{SIIIa} \\
  &\cS^{(\III,\y)}_{ik} = 
  \rS^{(\y)}_{ik} - (1\lra 2) - (3\lra 4) + \left(\substack{1\lra 3\\2\lra 4}\right) \label{SIIIb} \\
  &\cS^{(\III,\z)}_{ik} = \left.
  \rS^{(\x)}_{ik}\right|_{\sigma=+1} -2\rS^{(\y)}_{ik} + \rS^{(\z)}_{ik}
  + (1\lra 2) + (3\lra 4)
  + \left(\substack{1\lra 3\\2\lra 4}\right) \;, \label{SIIIc}
  \end{align}
where
\begin{subequations}\label{rS}
\begin{align}
  -\rS^{(\x)}_{ik}&(q_1,q_2,q_3,q_4;\sigma) =
  (q_{1234}^2\per q_{34}^2 )^{-2} \per (\pq{i}{1234}\per \pq{k}{1234})^{-1} 
  \Big\{ (q_{134}^2)^{-2} \nonumber \\
  &\quad\times \big[
   4\, \tr(p_i\,q_2\,p_k\,q_{134}\,q_4\,q_1\,q_3\,q_{134})
  +2 (d-2) \qq{3}{4}\,\tr(p_i\,q_2\,p_k\,q_{134}\,q_1\,q_{134})\big]
  \nonumber \\
  &\quad +\sigma\, (q_{134}^2\per q_{234}^2)^{-1}\big\{
   4\, \tr(p_i\,q_1\,q_4\,q_{134}\,p_k\,q_2\,q_3\,q_{234})
  \nonumber \\
  &\quad +2\, \qq{3}{4} \big[ (d-4)\tr(p_i\,q_1\,q_{134}\,p_k\,q_2\,q_{234})
 + 2\,\tr(p_i\,q_2\,q_1\,p_k\,q_{234}\,q_{134})
  \big]\big\} \Big\} \label{rSa} \\
  -\rS^{(\y)}_{ik}&(q_1,q_2,q_3,q_4) =
  [(q_{1234}^2)^2\per q_{12}^2 \per q_{34}^2\per q_{123}^2
  \per q_{134}^2 \per \pq{i}{1234}\per \pq{k}{1234}]^{-1} 
    \nonumber \\
  &\quad\big\{
  \tr(p_i\,q_2\,q_1\,q_3)\tr(p_k\,q_4\,q_{134}\,q_{123})
  +\tr(p_i\,q_2\,q_1\,q_4)\tr(p_k\,q_3\,q_{134}\,q_{123})
  \nonumber \\
  &\quad -\tr(p_i\,p_k\,q_1\,q_2)\tr(q_3\,q_4\,q_{123}\,q_{134})
  -\tr(p_i\,q_1\,p_k\,q_{134})\tr(q_2\,q_3\,q_4\,q_{123})
  \nonumber \\
  &\quad  -\tr(p_i\,q_2\,p_k\,q_{134})\tr(q_1\,q_3\,q_4\,q_{123})
  +\tr(p_i\,q_1\,q_3\,q_{134})\tr(p_k\,q_4\,q_2\,q_{123})
    \nonumber \\
  &\quad +{\textstyle\frac12} \big[ \tr(p_i\,q_2\,q_3\,q_{134})\tr(p_k\,q_4\,q_1\,q_{123})
  +\tr(p_i\,q_1\,q_4\,q_{134})\tr(p_k\,q_3\,q_2\,q_{123})
    \nonumber \\
  &\quad +\tr(p_i\,q_3\,q_4\,q_{123})\tr(p_k\,q_1\,q_2\,q_{134})
  +\tr(p_i\,q_2\,q_1\,q_{123})\tr(p_k\,q_4\,q_3\,q_{134})
    \nonumber \\
  &\quad +(d-4)\tr(p_i\,q_2\,q_1\,q_{134})\tr(p_k\,q_4\,q_3\,q_{123})
  \big] \big\} \label{rSb} \\
  -\rS^{(\z)}_{ik}&(q_1,q_2,q_3,q_4) =
  ( \pq{i}{1234}\per \pq{k}{1234} )^{-1} \bigg\{
  [q_{1234}^2 \per q_{12}^2\per (q_{34}^2)^2\per q_{234}^2 ]^{-1}
  \per 4 \left( \frac1{\pq{i}{12}}-\frac{1}{\pq{i}{34}}\right)
    \nonumber \\
 &\quad\times[2\,\pq{i}{4}\,\tr(p_k\,q_1\,p_i\,q_2\,q_3\,q_{234})
  -\qq{3}{4}\,\tr(p_i\,q_2\,p_i\,q_1\,p_k\,q_{234})]
  \nonumber \\
 &+[q_{1234}^2\per(q_{12}^2\per q_{34}^2)^2]^{-1} \,2
   \left( \frac1{\pq{i}{12}}-\frac{1}{\pq{i}{34}}\right)
  \big[\tr(p_i\,q_1\,p_k\,q_2)\tr(p_i\,q_3\,q_{12}\,q_4)
    \nonumber \\
 &\quad+\pq{k}{34} \big(4\pq{i}{1}\,\tr(p_i\,q_3\,q_2\,q_4)
  -2\,\qq{1}{2}\,\tr(p_i\,q_3\,p_i\,q_4) \big) \big] \nonumber \\
 &+[q_{12}^2\per q_{34}^2]^{-2} \frac18
   \left( \frac1{\pq{i}{12}}-\frac{1}{\pq{i}{34}}\right)
   \left( \frac1{\pq{k}{12}}-\frac{1}{\pq{k}{34}}\right)
    \tr(p_i\,q_1\,p_k\,q_2)\tr(p_i\,q_3\,p_k\,q_4) \nonumber \\
 & +[(q_{1234}^2 \per q_{34}^2)^2\per q_{12}^2\per q_{134}^2]^{-1}
 \nonumber \\
 &\quad \times 8\,\big\{4\,\qq{12}{3}\,\tr(p_i\,q_2\,p_k\,q_1\,q_4\,q_{134})
 -4\,\pq{k}{3}\,\tr(p_i\,q_2\,q_{34}\,q_1\,q_4\,q_{134})
    \nonumber \\
 &\quad +2\,\qq{3}{4}[\tr(p_i\,q_2\,q_{34}\,q_1\,p_k\,q_{134})-\tr(p_i\,q_2\,p_k\,q_1\,q_{12}\,q_{134}) ]
    \nonumber \\
 &\quad +p_k (q_{34}-q_{12}) [2\,\tr(p_i\,q_2\,q_3\,q_1\,q_4\,q_{134})
  +(d-2)\qq{3}{4}\,\tr(p_i\,q_2\,q_1\,q_{134}) ]\big\} \nonumber \\
 & +[q_{1234}^2 \per q_{12}^2 \per q_{34}^2]^{-2}
     \nonumber \\
 &\quad \times\big\{ 4\,
  [\tr(p_i\,q_1\,p_k\,q_2)\tr(q_3\,q_{12}\,q_4\,q_{12})
  -\tr(p_i\,q_1\,q_{34}\,q_2)\tr(p_k\,q_3\,q_{12}\,q_4)]
    \nonumber \\
 &\quad +16\,p_i (q_{34}-q_{12}) [ 2\,\pq{k}{2}\,\tr(q_4\,q_1\,q_3\,q_{12})
  -\qq{1}{2}\,\tr(p_k\,q_3\,q_{12}\,q_4)]
    \nonumber \\
  &\quad +4\,p_i (q_{34}-q_{12})\,p_k(q_{34}-q_{12})
  [\tr(q_1\,q_3\,q_2\,q_4)+2(d-2)\qq{1}{2}\,\qq{3}{4}]
  \big\} \bigg\} \;.
 \label{rSc}
\end{align}
\end{subequations}
While the functions $\cS^{(\III,\x)}_{ik}$ and $\cS^{(\III,\z)}_{ik}$ are symmetric in any of the exchanges $(1\lra 2)$ or $(3\lra 4)$, the function $\cS^{(\III,\y)}_{ik}$ in Eq.~\eqref{SIIIb} is symmetric only in the simultaneous exchange $\left(\substack{1\lra 2\\3\lra 4}\right)$. All of them (and thus $W^{(\III)}$) are also symmetric in the exchange $\left(\substack{1\lra 3\\2\lra 4}\right)$.

In summary, the squared current for soft $\qqbar\QQbar$ emission is
obtained by summing the contributions of Eqs.~(\ref{Jsq1}), \eqref{W1}, \eqref{W2} and \eqref{W3}, and we find
\begin{align}\label{Jsq2}
  &|\bj(q_1,q_2,q_3,q_4)|^2
  \eqcs \sy{|\bj(q_1,q_2)|^2}{|\bj(q_3,q_4)|^2} + W(q_1,q_2,q_3,q_4) \;,
\end{align}
where
\begin{align}
  W &= W^{(\I)} +W^{(\II)} +W^{(\III)} \nonumber \\
  &\eqcs-\left( \g\,\mu^\ep\right)^8 T_R^2 \left\{\sum_{i,k} \T_i\cdot \T_k \left[
    C_A\,\cS^{(A)}_{ik} + C_F\,\cS^{(F)}_{ik} \right]
  + \sum_{i,k,m} \tri{ikm} \cS_{ikm}\right \} \;, \label{Wqcd} \\
  \cS^{(A)}_{ik} &= \cS^{(\I)}_{ik} + \cS^{(\II)}_{ik}
  +\frac{1}{4} \left( -\cS^{(\III,\x)}_{ik} -6 \cS^{(\III,\y)}_{ik} +\cS^{(\III,\z)}_{ik} \right)
  \;,\quad
  \cS^{(F)}_{ik} = \cS^{(\III,\x)}_{ik} + 4 \cS^{(\III,\y)}_{ik} \;. \label{SikAF}
\end{align}
All dipole coefficients are charge-symmetric, i.e., symmetric in the exchange of both quark
and antiquark momenta $\left(\substack{1\lra 2\\3\lra 4}\right)$,
while the coefficient $\cS_{ikm}$ of the tripole features charge asymmetry, i.e., it is antisymmetric in such exchange.

The kinematical functions $\cS^{(A)}_{ik}$, $\cS^{(F)}_{ik}$ and $\cS_{ikm}$ in
Eq.~\eqref{Wqcd} depend on the hard-parton momenta (through the indices $i,k,m$), and we recall
that our results are valid for both massless and massive hard
partons. We note that the dipole kinematical functions $\cS^{(A)}_{ik}$ and $\cS^{(F)}_{ik}$
have an explicit dependence on the number $d=4-2\epsilon$ of space-time dimensions
[cfr.\ Eqs.~\eqref{rS}]. Such $\epsilon$ dependence actually derives by assuming (CDR) standard anticommutation relations between the Dirac gamma matrices:
$\{\gamma^\mu,\gamma^\nu\}=2g^{\mu\nu}$ so that $\gamma^\mu\gamma_\mu=g^\mu\null_\mu=d$,
$\gamma^\mu\gamma^\alpha\gamma_\mu=(2-d)\gamma^\alpha$ and so on.
In other versions of dimensional regularizations, such as dimensional reduction (DR) \cite{Siegel:1979wq} and the four-dimensional helicity (4DH) scheme \cite{Bern:1991aq},
the result for $|\bj(q_1,q_2,q_3,q_4)|^2$ is obtained by simply setting $\epsilon=0$
(i.e., $d=4$) in our expressions for $\cS$ and $\rS$.

We note that the contributions to Eq.~(\ref{Wqcd}) that are proportional to 
$\cS^{(A)}_{ik}$ have a purely non-abelian origin. The other contributions to
$W$ in Eq.~(\ref{Wqcd}), namely, the dipole terms proportional to $\cS^{(F)}_{ik}$
and the tripole terms, have an `abelian character', since corresponding irreducible correlations also occur for soft photon-lepton-antilepton radiation in QED
(see Sect.~\ref{s:qed}).

The structure of Eqs.~(\ref{Jsq2}) and (\ref{Wqcd}) is identical to that obtained in 
Ref.~\cite{Chen:2024hvp}. Exploiting colour conservation and the symmetries of
$\T_i\cdot \T_k$ and $\tri{ikm}$ with respect to their parton indices, the kinematical
functions $\cS^{(A)}_{ik}, \cS^{(F)}_{ik}$ and $\cS_{ikm}$ can be written in different ways,
without affecting the effect of $W(q_1,q_2,q_3,q_4)$ in Eq.~(\ref{Wqcd}) onto colour singlet states (scattering amplitudes). Our explicit expressions of these kinematical functions appear to be more compact than the related expressions presented in 
Ref.~\cite{Chen:2024hvp}. Considering the action of $W$ onto colour singlet states, we have carried out numerical comparisons between our result and the result of 
Ref.~\cite{Chen:2024hvp}. We find agreement for the terms with the symmetric product of uncorrelated $\qqbar$ squared currents and also for the tripole contribution, but we find disagreement for the dipole-correlated terms.

Following the suggestions of Ref.~\cite{Catani:2022hkb}, the irreducible correlation  $W(q_1,q_2,q_3,q_4)$ in Eq.~(\ref{Wqcd}) can be rewritten in the following equivalent form:
\begin{align}
  W(q_1\cdots q_4) 
  &\eqcs-\left( \g\,\mu^\ep\right)^8 T_R^2  \;\Bigl\{ \;\frac{1}{2} \sum_{i \neq k}
 \T_i\cdot \T_k \left[
    C_A\,w^{(A)}_{ik}(q_1\cdots q_4) + C_F\,w^{(F)}_{ik}(q_1\cdots q_4) \right] \Bigr.
  \nn \\
  & \Bigl.  + \frac{1}{2} \sum_{i \neq k} \tri{iik}\;w^{(tri)}_{ik}(q_1\cdots q_4)
+ \sum_{{\rm dist.}\{i,k,m\}} \tri{ikm} \;w^{(tri)}_{ikm}(q_1\cdots q_4) \;\Bigr\} \;,
\label{Wred}
\end{align}
where the subscript `${\rm dist.}\{i,k,m\}$' in $\sum_{{\rm dist.}\{i,k,m\}}$
denotes the sum over distinct hard-parton indices $i,k$ and $m$
(i.e., $i \neq k, k\neq m, m\neq i$).
The dipole kinematical functions $w^{(A)}_{ik}$ and $w^{(F)}_{ik}$ are related to the corresponding functions $\cS^{(A)}_{ik}$ and $\cS^{(F)}_{ik}$ in Eq.~(\ref{Wqcd}),
and we have
\begin{equation}\label{wrik}
  w^{(r)}_{ik}(q_1\cdots q_4) = \cS^{(r)}_{ik}(q_1\cdots q_4) + \cS^{(r)}_{ki}(q_1\cdots q_4)
  - \cS^{(r)}_{ii}(q_1\cdots q_4) - \cS^{(r)}_{kk}(q_1\cdots q_4) \,, \;
(r=A,F)\,.
\end{equation}
The kinematical functions $w^{(tri)}_{ik}$ and $w^{(tri)}_{ikm}$ depend on the tripole
functions $\cS_{ikm}$ in Eq.~(\ref{Wqcd}), and we obtain
\begin{align}
  w^{(tri)}_{ik}(q_1\cdots q_4) &= \Bigl[  \cS_{iik}(q_1\cdots q_4) + \cS_{iki}(q_1\cdots q_4)
  +  \cS_{kii}(q_1\cdots q_4) - \cS_{iii}(q_1\cdots q_4) \Bigr] \nn \\
 & - \;( i \leftrightarrow k) \;, \label{w3ik} \\
  w^{(tri)}_{ikm}(q_1\cdots q_4) &= \cS_{ikm}(q_1\cdots q_4) - \frac{1}{2} \Bigl[  \cS_{iik}(q_1\cdots q_4) + \cS_{iki}(q_1\cdots q_4) \Bigr. \nn \\ 
 & + \Bigl. \cS_{kii}(q_1\cdots q_4) - \cS_{iii}(q_1\cdots q_4) \Bigr] \;. \label{w3ikm}
\end{align}
The equality of Eqs.~(\ref{Wqcd}) and (\ref{Wred}) follows from colour conservation and the symmetries of the operators  $\T_i\cdot \T_k$ and $\tri{ikm}$ with respect to their 
hard-parton indices.

We note \cite{Catani:2021kcy} that the charge asymmetry contributions of
$|\bj(q_1,q_2,q_3,q_4)|^2$ vanish if the squared current acts on a pure multigluon scattering
amplitude $\M(\{p_i\})$, namely, if $\M(\{p_i\})$ has only gluon external lines (with no additional
external $\qqbar$ pairs or colourless particles). We also note \cite{Catani:2021kcy}
that the three-particle correlations of the type $\tri{ikm}$ with three distinct partons
contribute only to processes with {\it four} or {\it more} hard partons.
General properties of the colour algebra of the $d$-type tripoles $\tri{ikm}$
and their action onto two and three hard-parton states are discussed in 
Ref.~\cite{Catani:2021kcy} (see also the Appendix of Ref.~\cite{Catani:2022hkb}).

\section{Processes with two and three hard partons\label{s:e23h}}

In this section we present the explicit expressions of $| \J(q_1,q_2,q_3,q_4)|^2$
for $\qqbar\QQbar$ emission from amplitudes with two and three hard partons.

\subsection{Soft $\qqbar\QQbar$ emission from two hard partons\label{s:e2h}}

We consider a generic scattering amplitude $\M_{BC}(\{p_i\})$ whose
external legs are two hard partons (denoted as $B$ and $C$), with momenta $p_B$ and $p_C$,
and additional colourless particles.
The two hard partons can be either a $\hq{\bar\hq}$
pair%
\footnote{We denote the hard (anti)quarks with boldface italic font, in order to distinguish
them from the soft (anti)quarks $\qqbar$ and $\QQbar$ and from the momenta $q_i$ of the soft particles.}
($\{ BC \} = \{ \hq{\bar\hq} \}$) (note that we specify $B=\hq$ and $C={\bar\hq}$)
or two gluons ($\{ BC \} = \{ gg \}$). There is only {\em one} colour singlet configuration
of the two hard partons, and the corresponding one-dimensional colour space is generated by a single colour state vector that we denote as $\ket{B C}$. The colour space amplitude
$\ket{\M_{BC}}$ is a colour singlet state and, therefore, it is directly proportional to 
$\ket{B C}$.
The squared current $|\J(q_1,\cdots,q_N) |^2$ in Eq.~(\ref{softsquared})
conserves the colour charge of the hard partons and, consequently, the state
$|\J|^2 \,\ket{B C}$ is also proportional to $\ket{B C}$, with c-number eigenvalue
$|\J(q_1,q_2,q_3,q_4) |^{2}_{\; BC}$ given by
\begin{align}
  |\J(q_1,q_2,q_3,q_4) |^{2}_{\; BC}&= \left( \g\,\mu^\ep\right)^8 T_R^2 \,C_F
  \Bigl\{ \,\Bigl[ C_F  \left( w_{BC}(q_1,q_2) \;w_{BC}(q_3,q_4) + w^{(F)}_{BC}(q_1,q_2,q_3,q_4) \right) \Bigr. \Bigr. \nn \\
 & + \Bigl. \Bigl. C_A \;w^{(A)}_{BC}(q_1,q_2,q_3,q_4) \Bigr] 
  + \frac{1}{2} \,d_A \,w^{(tri)}_{BC}(q_1,q_2,q_3,q_4)\Bigr\} \;\;, \;\;
  ( B=\hq, C={\bar \hq}) \label{BCqq}
\end{align}
for $\ket{B C}=\ket{\hq \bar{\hq}}$ and by
\begin{align}
  | \J(q_1,q_2,q_3,q_4) |^{2}_{\; BC}&=\left( \g\,\mu^\ep\right)^8 T_R^2 \,C_A
  \,\Bigl\{ \; C_A  \left( w_{BC}(q_1,q_2) \;w_{BC}(q_3,q_4) + w^{(A)}_{BC}(q_1,q_2,q_3,q_4) \right) \Bigr.  \nn \\
& +  C_F \;w^{(F)}_{BC}(q_1,q_2,q_3,q_4) 
\Bigr\} \;\;, \quad \quad \quad \quad \quad \quad \quad \quad \quad \quad \quad 
 ( \{B C\}=\{gg\}) \;,
\label{BCgg}
\end{align}
for $\ket{B C}=\ket{gg}$. The colour coefficient $d_A$ is related to $d^{abc}$ as follows
\begin{equation}
  \sum_{bc} \,d^{abc} \,d^{dbc} = d_A \;\delta^{ad} \;\;, \quad\quad
  d_A= \frac{N_c^2-4}{N_c} \;.
\end{equation}
The dipole kinematical functions 
$w_{BC}(q_a,q_b)$, $w^{(A)}_{BC}(q_1,q_2,q_3,q_4)$ and $w^{(F)}_{BC}(q_1,q_2,q_3,q_4)$
are given in Eqs.~(\ref{w12}) and (\ref{wrik}),
and they are {\it symmetric} under the exchange $B \leftrightarrow C$ of the hard partons
(i.e., the exchange $p_B \leftrightarrow p_C$ of the hard-parton momenta).
The tripole kinematical function $w^{(tri)}_{BC}(q_1,q_2,q_3,q_4)$ is given in Eq.~(\ref{w3ik}), and it is {\it antisymmetric} under the exchange $B \leftrightarrow C$.

In the case of soft $\qqbar\QQbar$ emission from the hard partons $\{BC\}=\{\hq{\bar \hq}\}$,
the square-bracket contribution in the right-hand side of Eq.~(\ref{BCqq}) is 
symmetric with respect to the exchange
$\left(\substack{1\lra 2\\3\lra 4}\right)$ of the momenta of the soft
quarks and antiquarks. The kinematical function $w^{(tri)}_{BC}(q_1,q_2,q_3,q_4)$ is instead
antisymmetric under such exchange and, therefore, the result
in Eq.~(\ref{BCqq}) explicitly shows the presence of charge asymmetry effects.
Since the function $w^{(tri)}_{BC}(q_1,q_2,q_3,q_4)$ is antisymmetric with respect to the separate exchanges $\left(\substack{1\lra 2\\3\lra 4}\right)$ and $(B \lra C)$, in Eq.~(\ref{BCqq})
the asymmetry in the momenta of the soft quark and antiquark is correlated with a corresponding asymmetry in the momenta $p_B$ and $p_C$ of the hard quark and antiquark.
In particular, $|\J(q_1,q_2,q_3,q_4)|^{2}_{\; BC}$ is invariant under the overall
exchange of fermions and antifermions (i.e., $\{q_1,q_3,p_B\}\leftrightarrow\{q_2,q_4,p_C\}$),
consistently with charge conjugation invariance.

In the case of soft $\qqbar\QQbar$ radiation from two hard gluons, the tree-level result
in Eq.~(\ref{BCgg}) shows no charge asymmetry effects. As argued in 
Ref.~\cite{Catani:2021kcy}
on the basis of charge conjugation invariance, the absence of charge asymmetry effects
in $|\J(q_1,q_2,q_3,q_4)|^{2}_{\; BC}$ for $\{BC\}=\{gg\}$ persists at arbitrary orders
in the QCD loop expansion.

\subsection{Soft $\qqbar\QQbar$ emission from three hard partons\label{s:e3h}}

We consider a generic scattering amplitude $\M_{ABC}(\{p_i\})$ whose
external legs are three hard partons and additional colourless particles.
The three hard partons, which are denoted as $A, B, C$ (with momenta $p_A, p_B ,p_C$),
can be either a gluon and a $\hq{\bar \hq}$ pair ($\{A B C\}=\{g \hq {\bar \hq}\}$)
or three gluons ($\{A B C\}=\{g g g\}$).

In the case $\{A B C\}=\{g \hq {\bar \hq}\}$, there is only one colour singlet state that can be formed by the three hard partons, hence the colour-singlet space is one-dimensional. Therefore, for the same reasoning as in Sect.~\ref{s:e2h}, 
the state $\ket{A B C}$ is an eigenstate of the squared current $|\J(q_1,\cdots,q_4)|^2$
in Eq.~(\ref{softsquared}), with $c$-number eigenvalue $|\J(q_1,\cdots,q_4)|^{2}_{\; ABC}$ given by 
\begin{align}
 |\J(q_1,q_2,q_3,q_4)|^{2}_{\; ABC} &= \left( \g\,\mu^\ep\right)^8 \,T_R^2
 \Bigl[ \, F_{ABC}^{({\rm in.em.})}(q_1,q_2,q_3,q_4) 
 + W_{ABC}^{({\rm ch.sym.})}(q_1,q_2,q_3,q_4) \Bigr. \nn \\
 &+ \Bigl. W_{ABC}^{({\rm ch.asym.})}(q_1,q_2,q_3,q_4) \,\Bigr] \;, 
 \quad \quad \quad \quad (\{A B C\}=\{g \hq {\bar \hq}\}) \;,
 \label{softgqq}
\end{align}
which directly derives from Eqs.~(\ref{Jsq2}) and (\ref{Wred}).
The functions $F_{ABC}^{({\rm in.em.})}$ and $W_{ABC}^{({\rm ch.sym.})}$ 
in Eq.~(\ref{softgqq}) are
\begin{align}
  F_{ABC}^{({\rm in.em.})}(q_1,q_2,q_3,q_4)
 &= \bigl[ C_F \;w_{BC}(q_1,q_2) + C_A \;w_{ABC}(q_1,q_2) \bigr] \nonumber \\
 & \times \bigl[ C_F \;w_{BC}(q_3,q_4) + C_A \;w_{ABC}(q_3,q_4) \bigr] \;,
 \quad (\{A B C\}=\{g \hq {\bar \hq}\}) \;, \label{fgqq} \\
  W_{ABC}^{({\rm ch.sym.})}(q_1 \cdots q_4) &= C_F^2 \;w_{BC}^{(F)}(q_1 \cdots q_4)  
  + C_F C_A \Big[ w_{BC}^{(A)}(q_1 \cdots q_4) + w_{ABC}^{(F)}(q_1 \cdots q_4) \Big] \nonumber \\
 &+ C_A^2 \;w_{ABC}^{(A)}(q_1 \cdots q_4) \;,
  \quad (\{A B C\}=\{g \hq {\bar \hq}\}) \;, \label{csgqq}
\end{align}
where the two hard-parton functions
$w_{ik}(q_a,q_b)$, $w^{(r)}_{ik}(q_1,q_2,q_3,q_4) \;(r=F,A)$ 
are given in Eqs.~(\ref{w12}) and (\ref{wrik}),
and we have used them to define the corresponding three hard-parton functions $w_{ABC}^{(r)}$:
\footnote{In the case of two soft momenta $(N=2)$, the explicit superscripts $(r)$ have to be removed in Eq.~(\ref{wabc}).}
\begin{equation}\label{wabc}
  w_{ABC}^{(r)}(q_1,\cdots,q_N) \equiv 
  \frac{1}{2} \left[ w_{AB}^{(r)}(q_1,\cdots,q_N) + 
    w_{AC}^{(r)}(q_1,\cdots,q_N) - w_{BC}^{(r)}(q_1,\cdots,q_N) 
    \right] \;.
\end{equation}
The function $W_{ABC}^{({\rm ch.asym.})}$ in Eq.~(\ref{softgqq}) is
\begin{align}
 W_{ABC}^{({\rm ch.asym.})}(q_1 \cdots q_4) &= \frac{d_A}{4}
 \Big[ - C_A \, \Big( w^{(tri)}_{AB}(q_1 \cdots q_4) + w^{(tri)}_{BC}(q_1 \cdots q_4)
 + w^{(tri)}_{CA}(q_1 \cdots q_4) \Big)\nonumber \\
  & + 2 C_F \, w^{(tri)}_{BC}(q_1 \cdots q_4) 
 \Big] \;, \quad (\{A B C\}=\{g \hq {\bar \hq}\}) \;,
\label{cagqq}
\end{align}
where the kinematical function $w^{(tri)}_{ik}$ is given in Eq.~(\ref{w3ik}).
As discussed below, the symmetry properties of the kinematical functions $w_{ik}(q_a,q_b)$, $w^{(r)}_{ik}(q_1\dots q_4)$ and 
$w^{(tri)}_{ik}(q_1\dots q_4)$ with respect to their dependence 
on the hard and soft momenta (see Sect.~\ref{s:qqQQsc})
lead to ensuing symmetry properties of the functions
$F^{({\rm in.em.})}$, $W^{({\rm ch.sym.})}$ and $W^{({\rm ch.asym.})}$.

The term $F^{({\rm in.em.})}_{ABC}$ in Eq.~(\ref{softgqq}) is the contribution due to the independent emission of the soft $\qqbar$ and the soft $\QQbar$ pair. It originates from the
term $\sy{|\bj(q_1,q_2)|^2}{|\bj(q_3,q_4)|^2}$ in Eq.~(\ref{Jsq2}). We have used the results of the action of both $|\bj(q_1,q_2)|^2$ and $|\bj(q_3,q_4)|^2$ onto the three hard-parton state
$\ket{ABC}$ \cite{Catani:1999ss}. Such squared currents only depend on colour dipole operators, whose action onto $\ket{ABC}$ is simply given in terms of Casimir coefficients 
$C_F$ and $C_A$ (see, e.g., Ref.~\cite{csdip}). 
The colour dipole contributions of Eq.~(\ref{Wred}) to the correlation term
$W(q_1,q_2,q_3,q_4)$
produce the corresponding
irreducible correlation contribution $W^{({\rm ch.sym.})}_{ABC}$ in Eq.~(\ref{softgqq}).
We note that both $F^{({\rm in.em.})}_{ABC}(q_1,q_2,q_3,q_4)$ and 
$W^{({\rm ch.sym.})}_{ABC}(q_1,q_2,q_3,q_4)$ are symmetric under the exchange
$\left(\substack{1\lra 2\\3\lra 4}\right)$ (see Eqs.~(\ref{fgqq}) and (\ref{csgqq})) and, therefore, they do not lead to any charge asymmetry of the soft quark and antiquark in 
$| \J(q_1,q_2,q_3,q_4) |^{2}_{\; ABC}$. Both functions $F^{({\rm in.em.})}_{ABC}$ and
$W^{({\rm ch.sym.})}_{ABC}$ are also symmetric under the exchange $p_B \leftrightarrow p_C$
of the momenta of the hard quark and antiquark, consistently with the charge conjugation
invariance of  $| \J(q_1,q_2,q_3,q_4) |^{2}_{\; ABC}$.

The correlation term $W(q_1,q_2,q_3,q_4)$ of Eq.~(\ref{Jsq2}) also includes charge asymmetry contributions. In Eq.~(\ref{Wred}) these contributions are proportional to the tripole
operators $\tri{iik}$ and $\tri{ikm}$. The action of these tripole operators onto the state
$\ket{ABC}$ of the three hard partons $\{A B C\}=\{g \hq {\bar \hq}\}$ was evaluated in 
Ref.~\cite{Catani:2021kcy} (in particular, the operator $\tri{ABC}$ with three distinct indices vanishes). Using the colour algebra results of Ref.~\cite{Catani:2021kcy}
(see also the Appendix of Ref.~\cite{Catani:2022hkb}), we have computed the charge asymmetry contribution to 
$| \J(q_1,q_2,q_3,q_4) |^{2}_{\; ABC}$, which is given by the function 
$W_{ABC}^{({\rm ch.asym.})}(q_1,q_2,q_3,q_4)$ in Eq.~(\ref{softgqq}).
We note the the expression of  $W_{ABC}^{({\rm ch.asym.})}$ in Eq.~(\ref{cagqq})
is antisymmetric under the exchange $p_B \leftrightarrow p_C$ of the momenta of the hard quark and antiquark, in complete analogy with the charge asymmetry contribution to 
Eq.~(\ref{BCqq}),
and consistently with the charge conjugation invariance of 
$| \J(q_1,q_2,q_3,q_4) |^{2}_{\; ABC}$ in Eq.~(\ref{softgqq}).

We now consider the case $\{A B C\}=\{g g g \}$. The three hard gluons generate a 
two-dimensional colour singlet space. We choose the basis that is formed by the two
colour state vectors $\ket{(ABC)_f \,}$ and
$\ket{(ABC)_d \,}$, which are defined as follows
\beq
\label{ABCbasis}
\bra{\,abc\,} \left(ABC\right)_f \,\rangle \equiv i f^{abc}
\,,
\;\;\;
\bra{\,abc\,} \left(ABC\right)_d \,\rangle \equiv  d^{abc}
\,,
\;\;\; \quad (\{ ABC \} = \{ ggg \}) \;\;,
\eeq
where $a, b, c$ are the colour indices of the three gluons. We note that the two states
in Eq.~(\ref{ABCbasis}) are orthogonal and have different charge conjugation.
The scattering amplitude $\ket{\cm_{ABC}(\{p_i\})}$ is, in general, a linear combination 
of the two states in Eq.~(\ref{ABCbasis}), and the action of the squared current 
$| \bj(q_1, \cdots, q_N) |^2 $ for soft-parton radiation onto $\ket{\cm_{ABC}(\{p_i\})}$
can produce colour correlations between these two states.
In general, $| \bj(q_1, \dots, q_N) |^2 $
can be represented as a $2 \times 2$ correlation matrix that
acts on the two-dimensional space generated by $\ket{\left( ABC \right)_f}$ and $\ket{\left( ABC \right)_d}$.
The structure of this correlation matrix is discussed in Refs.~\cite{Catani:2019nqv}, \cite{Catani:2022hkb} and \cite{Catani:2021kcy}
for the cases of multiple soft-gluon radiation, soft-$\qqbar$ radiation and soft $g\qqbar$ radiation,
respectively. Soft $\qqbar\QQbar$ radiation is discussed in the following.

The action of the tree-level squared current $| \J(q_1,q_2,q_3,q_4) |^{2}$ for 
soft $\qqbar\QQbar$ emission onto the colour singlet states in Eq.~(\ref{ABCbasis})
can be written in the following form:
\begin{align}
| \J(q_1,q_2,q_3,q_4) |^{2} \;\ket{ABC} &= \left( \g\,\mu^\ep\right)^8 \,T_R^2
\Bigl[ \, F_{ABC}^{({\rm in.em.})}(q_1,q_2,q_3,q_4) 
+ W_{ABC}^{({\rm ch.sym.})}(q_1,q_2,q_3,q_4) \Bigr. \nn \\
&+ \Bigl. W_{ABC}^{({\rm ch.asym.})}(q_1,q_2,q_3,q_4) \,\Bigr] \;\ket{ABC}\;, 
\quad \quad \quad \quad (\{A B C\}=\{ggg\}) \;,
\label{J2ggg}
\end{align}
where $F_{ABC}^{({\rm in.em.})}$ and $W_{ABC}^{({\rm ch.sym.})}$ are c-number functions
\begin{equation}
\label{inem3g}
F_{ABC}^{({\rm in.em.})}(q_1,q_2,q_3,q_4) = 4 C_A^2 \;E_{ABC}(q_1,q_2) \;E_{ABC}(q_3,q_4) \;\;,
\quad \quad (\{A B C\}=\{ggg\}) \;,
\end{equation}
\begin{align}
W_{ABC}^{({\rm ch.sym.})}(q_1,q_2,q_3,q_4) = 2 C_A 
\!\left[ C_F \,E_{ABC}^{(F)}(q_1,q_2,q_3,q_4) +  C_A \,E_{ABC}^{(A)}(q_1,q_2,q_3,q_4) \right]\;,
 \nonumber \\ \hfill (\{A B C\}=\{ggg\}) \;, \label{symggg}
\end{align}
while
\begin{equation}
\label{asymggg}
W_{ABC}^{({\rm ch.asym.})}(q_1,q_2,q_3,q_4) = 4 \;E_{ABC}^{(tri)}(q_1,q_2,q_3,q_4) \;\tri{BBA} \;\;,
\quad \quad \;\;\;\; (\{A B C\}=\{ggg\})
\end{equation}
contains the tripole operator $\tri{BBA}$ acting non-trivially onto the colour-singlet state $\ket{ABC}$ (see Eq.~\eqref{triggg} below).
The three hard-parton functions $E_{ABC}(q_a,q_b)$ and 
$E_{ABC}^{(r)}(q_1,q_2,q_3,q_4)$ (with $r=F,A,tri$) in Eqs.~(\ref{inem3g})--(\ref{asymggg})
are given as follows:
\footnote{In the case of two soft momenta $(N=2)$, the explicit superscripts $(r)$ have to be removed in Eq.~(\ref{eabc}).}
\begin{equation}
\label{eabc}
  E_{ABC}^{(r)}(q_1,\cdots,q_N) \equiv 
  \frac{1}{4} \left[ w_{AB}^{(r)}(q_1,\cdots,q_N) + 
    w_{BC}^{(r)}(q_1,\cdots,q_N) + w_{CA}^{(r)}(q_1,\cdots,q_N) 
    \right] \;,
\end{equation}
in terms of the corresponding two hard-parton functions
$w_{ik}(q_a,q_b)$, $w^{(r)}_{ik}(q_1,q_2,q_3,q_4)$ 
in Eqs.~\eqref{w12}, \eqref{wrik} and \eqref{w3ik}.
The symmetry properties of the functions $E_{ABC}^{(r)}$
are the consequence
of the corresponding symmetries of the functions 
$w_{ik}^{(r)}$
in the right-hand side of Eq.~(\ref{eabc}).
The function $E_{ABC}^{(tri)}(q_1,q_2,q_3,q_4)$ is antisymmetric under the exchange 
$\left(\substack{1\lra 2\\3\lra 4}\right)$ of the momenta of the soft quarks and antiquarks, and it is also antisymmetric under the exchange of the momenta of two hard gluons (e.g., 
$p_A \leftrightarrow p_B$). The functions $E_{ABC}^{(r)}$ in Eqs.~(\ref{inem3g}) and 
(\ref{symggg}) are instead symmetric under the exchange 
$\left(\substack{1\lra 2\\3\lra 4}\right)$ and have a fully symmetric dependence on the hard-gluon momenta
$p_A, p_B, p_C$.

The result in Eq.~(\ref{J2ggg}) directly derives from Eqs.~(\ref{Jsq2}) and (\ref{Wred}),
and it has a structure that follows the structure of Eq.~(\ref{softgqq}).
The term $F^{({\rm in.em.})}_{ABC}(q_1,q_2,q_3,q_4)$ in Eq.~(\ref{J2ggg}) is the contribution
of the independent emission of the soft $\qqbar$ pair and the soft $\QQbar$ pair. 
The irreducible correlation term $W^{({\rm ch.sym.})}_{ABC}(q_1,q_2,q_3,q_4)$
in Eq.~(\ref{J2ggg}) is due to the colour dipole contributions of Eq.~(\ref{Wred}) to 
$W(q_1,q_2,q_3,q_4)$. Both terms $F^{({\rm in.em.})}_{ABC}$ and $W^{({\rm ch.sym.})}_{ABC}$
originate from colour dipole interactions, whose action onto a generic hard-parton state
$\ket{ABC}$ are simply proportional to the unit operator in colour space 
\cite{Catani:1999ss}.

The term $W_{ABC}^{({\rm ch.asym.})}$ in Eq.~(\ref{J2ggg}) is due to the colour tripole
contributions of Eq.~(\ref{Wred}) to the irreducible correlation operator $W(q_1,q_2,q_3,q_4)$
The action of the tripole operators onto the three-gluon states in Eq.~(\ref{ABCbasis}) was explicitly evaluated in Ref.~\cite{Catani:2021kcy} (see also Appendix of Ref.~\cite{Catani:2022hkb}). In particular, the tripoles $\tri{ABC}$ with three distinct gluons vanish, while the tripoles with two distinct gluons are proportional to one another
(the proportionality factors are $\pm 1$). It turns out that the term $W_{ABC}^{({\rm ch.asym.})}$ in Eq.~(\ref{J2ggg}) is directly proportional to a single tripole operator
(e.g., $\tri{BBA}$) as shown in Eq.~(\ref{asymggg}). The tripole operators are odd under charge conjugation and, therefore, they act differently onto the two colour states in 
Eq.~(\ref{ABCbasis}). Considering the operator $\tri{BBA}$ in Eq.~(\ref{asymggg}), we have
\cite{Catani:2021kcy}
\beq
\label{triggg}
\tri{BBA} \,\ket{\left( ABC \right)_f} =  \frac{C_A^2}{4}  \;\ket{\left( ABC \right)_d} \;,
\;\;\;\;
\tri{BBA} \,\ket{\left( ABC \right)_d} = \frac{C_A \,d_A}{4}  \;\ket{\left( ABC \right)_f} \;,
\eeq
and we note that the tripole operators produce 
`pure' 
transitions between the colour 
symmetric and colour antisymmetric states $\ket{\left( ABC \right)_f}$ and $\ket{\left( ABC \right)_d}$,
which have different charge conjugation.

The results in Eqs.~(\ref{J2ggg})--(\ref{asymggg}) can be used to explicitly evaluate the action of the tree-level squared current $| \J(q_1,q_2,q_3,q_4) |^{2}$ onto a scattering amplitude $\ket{\cm_{ABC}(\{p_i\})}$ with three hard gluons.
The 
scattering amplitude $\ket{\cm_{ABC}(\{p_i\})}$ is, in general, a linear combination 
of the two colour states in Eq.~(\ref{ABCbasis}), and we write
\beq
\label{3gamp}
\ket{\cm_{ABC}(\{p_i\})} = \ket{\left( ABC \right)_f} \;\;\cm_{f}(p_A,p_B,p_C) \,+
\ket{\left( ABC \right)_d} \;\;\cm_{d}(p_A,p_B,p_C) \;\;, 
\eeq
where $\cm_f$ and $\cm_d$ are colour-stripped amplitudes.
Owing to the Bose symmetry of $\ket{\cm_{ABC}}$ with respect to the three gluons,
the function $\cm_{f}(p_A,p_B,p_C)$ is antisymmetric under the exchange of two gluon momenta
(e.g., $p_A \leftrightarrow p_B$), while $\cm_{d}(p_A,p_B,p_C)$ has a symmetric dependence on
$p_A,p_B,p_C$.
Using Eqs.~(\ref{J2ggg})--(\ref{asymggg}), (\ref{triggg}) and (\ref{3gamp})
we obtain
\begin{align}
&\bra{\M_{ABC} (\{p_i\})} \;| \J(q_1,q_2,q_3,q_4) |^2 \;\ket{\M_{ABC} (\{p_i\})} 
= \left( \g\,\mu^\ep\right)^8 \,T_R^2 \nn \\
&\;\;\;\; \times \Bigl\{ | \M_{ABC} (\{p_i\}) |^2 \; 
\Bigl[ \, F_{ABC}^{({\rm in.em.})}(q_1,q_2,q_3,q_4) 
+ W_{ABC}^{({\rm ch.sym.})}(q_1,q_2,q_3,q_4) \Bigr] \Bigr. \nn \\
&\;\;\;\;\;\;\; + \Bigl. C_A^2 d_A (N_c^2-1) 
\bigl[ \, \cm^\dagger_d(p_A,p_B,p_C)  \cm_f(p_A,p_B,p_C) + {\rm h.c.} \bigr]
E_{ABC}^{(tri)}(q_1,q_2,q_3,q_4)
\Bigr\} \;, \nonumber \\
& \quad~~~~~~~~~~~~~~~~~~~~~~~~~~~ ~~~~~~~~~~~~~~~~~~~ ~~~~~~~~~~~~~~~~~~~~~~~~~~~\quad 
(\{A B C\}=\{ggg\}) \;, \label{m23g}
\end{align}
which is not simply proportional to $| \cm_{ABC}(\{p_i\}) |^2$ (unlike the corresponding result in Eq.~(\ref{softgqq}) for $\{ ABC \} = \{ g\hq{\bar\hq} \}$). 
The contribution to Eq.~(\ref{m23g}) that is proportional to 
$F_{ABC}^{({\rm in.em.})} + W_{ABC}^{({\rm ch.sym.})}$ is symmetric under the exchange
$\left(\substack{1\lra 2\\3\lra 4}\right)$ and, therefore, it does not lead to any charge asymmetry of the soft quark and antiquark. The function $E_{ABC}^{(tri)}(q_1,q_2,q_3,q_4)$ is instead 
antisymmetric under the exchange
$\left(\substack{1\lra 2\\3\lra 4}\right)$. Therefore, in contrast with the case of scattering amplitudes with two hard gluons (see Eq.~(\ref{BCgg})), the expression in Eq.~(\ref{m23g})
involves a charge-asymmetry contribution that is not vanishing,
provided the hard-scattering amplitude includes non-vanishing components $\cm_f$ and $\cm_d$
(i.e., $\ket{\cm_{ABC}}$ has no definite charge conjugation). This is the case, for instance,
of the amplitude for the decay process $Z \to ggg$ of the $Z$ boson
(see, e.g., Ref.~\cite{vanderBij:1988ac}).
We note that the functions $E_{ABC}^{(tri)}$ and $(\cm_d^\dagger \cm_f + {\rm h.c.})$ 
are separately antisymmetric under the exchange of two gluon momenta and, consequently, their
product is symmetric. Therefore, the right-hand side of Eq.~(\ref{m23g})
is fully symmetric under permutations of the three
hard gluons, as required by Bose symmetry.

\section{QED and mixed QCD$\times$QED interactions\label{s:qed}}

Our results of Sects.~\ref{s:qqQQc} and \ref{s:qqQQsc}
for soft $\qqbar\QQbar$ radiation at tree level in QCD are generalized in this section to deal with soft emission through QED (photon) interactions and mixed QCD$\times$QED (gluon and photon)
interactions.

We consider a generic scattering amplitude $\M(\{q_r\}, \{p_i\})$ whose external soft {\it massless} particles $\{q_r\}$ are fermions ($f$) and antifermions ($\bar f$).
The soft massless fermions are quarks ($f=\quark$) and charged leptons ($f=\ell$). The external massless and massive hard partons $\{p_i\}$ in $\M$ are gluons, (anti)quarks and electrically charged particles, such as (anti)leptons and $W^\pm$ bosons.
The amplitude $\M$ can also have external particles that carry no colour charge and no electric charge. 

We formally treat QCD, QED and mixed QCD$\times$QED interactions on equal footing. Therefore, the scattering amplitude $\M$ has a generalized perturbative (loop) expansion in powers of two unrenormalized couplings: the QCD coupling $\g$ and the QED coupling $\gq$ ($\gq^2/(4\pi)=\alpha$ is the fine structure constant at the unrenormalized level). In the soft limit
the amplitude $\M(\{q_r\}, \{p_i\})$ fulfils the factorization formula (\ref{1gfact}),
and the soft current $\J(q_1,\cdots,q_N)$ also has a loop expansion in powers of the two couplings $\g$ and $\gq$. In the following we only consider soft currents at tree level
with respect to both couplings and, therefore, the $N$ parton current 
$\J(q_1,\cdots,q_N)$ includes all possible contributions that are proportional to 
$\g^{N-k} \gq^k$ with $0 \leq k \leq N$. The pure QCD and pure QED cases are recovered by setting $\gq=0$ and $\g=0$, respectively.

We first recall the known expressions of the soft current for fermion-antifermion emission.
The tree-level current $\bj_{f{\bar f}}(q_1,q_2)$ for emission of a soft-$f{\bar f}$ pair is \cite{Catani:2021kcy}
\begin{align}\label{J1ff}
  \bj_{f{\bar f}}(q_1,q_2) &= \delta_{f\quark} \;\bj(q_1,q_2)
  - \left( \gq \,\mu^\ep \right)^2  \,e_f \;{\bf \Delta}_f
  \, \sum_{i} e_i \;\frac{p_i \cdot j(1,2)}{p_i \cdot q_{12}} \;,
\end{align}
where $q_1$ and $q_2$ are the momenta of the soft fermion $f$ and antifermion $\bar f$, respectively. The first contribution in the right-hand side of Eq.~(\ref{J1ff}) is the
QCD current $\bj(q_1,q_2)$ in Eq.~(\ref{Jqa})
(the Kronecker delta symbol $\delta_{f\quark}$ specifies that the current is not vanishing only
if the fermon $f$ is a quark), and the second contribution is due to the photon mediated radiation of the 
$f{\bar f}$ pair. The fermionic current $j^\nu(1,2)$ is given in Eq.~(\ref{fercur}),
and $e_f$ ($e_i$) is the electric charge of the soft fermion $f$ (of the hard $i$-th particle) in units of the positron charge.
The factor ${\bf \Delta}_f$ in the right-hand side of Eq.~(\ref{J1ff})
is a colour operator that depends on the type
of soft fermion $f$. If $f=\ell$, we simply have ${\bf \Delta}_f= 1$. If $f=\quark$, 
${\bf \Delta}_f$ is the projection operator onto the colour singlet
state of the $f{\bar f}$ pair, namely, by using the colour space notation of
Sect.~\ref{s:sfsc} we have
$\bra{\alpha_1,\alpha_2}  \,{\bf \Delta}_f = \delta_{\alpha_1\alpha_2}$, where
$\alpha_1$ and $\alpha_2$ are the colour indices of the soft quark and antiquark, respectively.

The square of the current $\bj_{f{\bar f}}(q_1,q_2)$ is \cite{Catani:2021kcy}
\begin{equation}\label{J2ff}
  |\bj_{f{\bar f}}(q_1,q_2)|^2 = \delta_{f\quark} \;|\bj(q_1,q_2)|^2
  - \left(\gq \,\mu^\ep \right)^4 \,\left( \delta_{f\ell}
  + N_c \,\delta_{f\quark} \right) \,e_f^2
  \sum_{i<k}  e_i\,e_k \;w_{ik}(q_1,q_2)\;\;,
\end{equation}
where $|\bj(q_1,q_2)|^2$ is the QCD squared current in Eq.~(\ref{j12q})
and the function $w_{ik}(q_1,q_2)$ is given in Eqs.~(\ref{I12}) and (\ref{w12}).
Similarly to its QCD part, the complete squared current $|\bj_{f{\bar f}}(q_1,q_2)|^2$
is charge symmetric with respect to the exchange $f \leftrightarrow {\bar f}$
(i.e., it is symmetric under $q_1 \leftrightarrow q_2$).
We note that the squared current result in Eq.~(\ref{J2ff}) does not include a mixed
QCD$\times$QED term proportional to $\g^2 \gq^2$, since such contribution vanishes.

In the following we present our results for radiation at tree level of two
distinguishable soft fermion-antifermion pairs $f{\bar f}F{\bar F}$, whose momenta
are denoted $q_1,q_2,q_3,q_4$ respectively.

\subsection{Soft emission of two lepton pairs\label{s:llLL}}

We first consider the case of radiation of two distinguishable soft lepton-antilepton
pairs $\llbar\LLbar$. The corresponding soft current can be written as
\begin{equation}\label{JllLL}
  \bj_{\llbar\LLbar}(q_1,q_2,q_3,q_4) =
  \bj^{\QED}_{\llbar}(q_1,q_2)\bj^{\QED}_{\LLbar}(q_3,q_4)
  + \Ga_{\llbar\LLbar}(q_1,q_2,q_3,q_4) \;,
\end{equation}
where the lepton-antilepton soft current is [cfr.~Eq.~\eqref{J1ff}]
\begin{align}\label{J1ll}
  \bj^{\QED}_{\llbar}(q_1,q_2) &= 
  - \left( \gq \,\mu^\ep \right)^2  \,e_\ell \, \sum_{i}
  e_i \;\frac{p_i \cdot j(1,2)}{p_i \cdot q_{12}}
\end{align}
and a similar expression for $\bj^{\QED}_{\LLbar}(q_3,q_4)$ with $(e_\ell\to e_L)$
besides $(1\to 3)$ and $(2\to 4)$.
At variance with the QCD current for soft quark pairs of Eq.~\eqref{J1234},
the symmetric product is not needed, because the two lepton-antilepton soft currents in the r.h.s.\ of Eq.~\eqref{JllLL} commute.

Since leptons interact only by photon exchange, only the diagrams
A, B, C and D of fig.~\ref{f:qqQQ} contribute to the irreducible correlation
\begin{subequations}\label{GllLL}
\begin{align}
  &\Ga_{\llbar\LLbar}(q_1,q_2,q_3,q_4) \eqcs \left(\gq\,\mu^\ep\right)^4 \,
  \sum_i e_i \,\sum_{\sigma\in\{\x,\y\}} \cols_{(\sigma)}^\QED \gamma_i^{(\sigma)}(q_1,q_2,q_3,q_4) \\
  &\cols_{(\x)}^\QED = e_\ell^2\,e_L \;,\quad
  \cols_{(\y)}^\QED = e_\ell\,e_L^2 \;.
\end{align}
\end{subequations}
Consequently, only the two ``abelian structures'' ($\sigma\in\{\x,\y\}$) contribute to the irreducible correlation $\Ga$ [cfr.~Eqs.~(\ref{G1234},\ref{R123})]; the kinematical coefficients
$\gamma_i^{(\sigma)}$ are the same of Eq.~\eqref{g12}.
Also in this case the dependence on the auxiliary gauge vector $n$ disappears
when the current $\bj_{\llbar\LLbar}$ acts on colour singlet states.
Therefore the expressions given above are gauge-independent.

The squared current for soft $\llbar\LLbar$ emission is computed by using the expressions in Eqs.~\eqref{JllLL} and \eqref{GllLL}. We write the result as follows
\begin{equation}\label{J2llLL}
  | \bj_{\llbar\LLbar}(q_1,q_2,q_3,q_4) |^2
  = |\bj^{\QED}_{\llbar}(q_1,q_2)|^2 \, |\bj^{\QED}_{\LLbar}(q_3,q_4)|^2
  + W_{\llbar\LLbar}(q_1,q_2,q_3,q_4) \;,
\end{equation}
where $|\bj^{\QED}_{\llbar}(q_1,q_2)|^2$ is the square of the current in Eq.~\eqref{J1ll}:
\begin{equation}\label{J2ll}
  |\bj^{\QED}_{\llbar}(q_1,q_2)|^2 = 
  \left(\gq \,\mu^\ep \right)^4 \,e_\ell^2\,
  \sum_{i,k}  e_i\,e_k \;{\cal I}_{ik}(q_1,q_2)
 = - \left(\gq \,\mu^\ep \right)^4 \,e_\ell^2
  \sum_{i < k}  e_i\,e_k \;w_{ik}(q_1,q_2)
\end{equation}
and similarly $|\bj^{\QED}_{\LLbar}(q_3,q_4)|^2$ by replacing $(1\to 3, 2\to 4, e_\ell \to e_L)$.

The irreducible correlation contribution $W_{\llbar\LLbar}$ reads%
\footnote{Note the exchange $1\lra 3,2\lra 4$ in the momenta of some $\cS$ functions.}
\begin{align}
W_{\llbar\LLbar} = -(\gq\, \mu^{\ep})^8 e_\ell^2\,e_L^2
  &\Big\{ 2\sum_{i,k,m} e_i e_k e_m
  \bigl[ e_\ell \cS^{(\ell)}_{ikm}(q_1,q_2,q_3,q_4)
   + e_L \cS^{(\ell)}_{ikm}(q_3,q_4,q_1,q_2)
  \bigr] \nonumber\\
 & + \sum_{i,k} e_i e_k \bigl[ e_\ell^2 \cS_{ik}^{(\III,\x,\ell)}(q_1,q_2,q_3,q_4)
  +e_L^2 \cS_{ik}^{(\III,\x,\ell)}(q_3,q_4,q_1,q_2) \nonumber \\
 &\qquad\qquad +e_\ell e_L \,2\cS_{ik}^{(\III,\y)}(q_1,q_2,q_3,q_4) \bigr]
 \Bigr\} \label{WllLL}
\end{align}
where the ``leptonic'' momentum dependent functions $\cS^{(\ell)}_{ikm}$ and
$\cS_{ik}^{(\III,\x,\ell)}$ are
\begin{align}
  \cS^{(\ell)}_{ikm}(q_1,q_2,q_3,q_4) &= 
  \rS_{ikm}(q_1,q_2,q_3,q_4) + (3\lra 4) - (1\lra 2) \\
  \cS^{(\III,\x,\ell)}_{ik}(q_1,q_2,q_3,q_4) &= \left.
  \rS^{(\x)}_{ik}(q_1,q_2,q_3,q_4)\right|_{\sigma=-1} + (1\lra 2) + (3\lra 4)
\end{align}
which are related to the functions $\cS_{ikm}$ and
$\cS^{(\III,\x)}_{ik}$ of Eqs.~\eqref{Sikm} and \eqref{SIIIa} as follows:
\begin{align}
  \cS_{ikm}(q_1,q_2,q_3,q_4)
  &= \cS^{(\ell)}_{ikm}(q_1,q_2,q_3,q_4) + \cS^{(\ell)}_{imk}(q_3,q_4,q_1,q_2) \\
  \cS^{(\III,\x)}_{ik}(q_1,q_2,q_3,q_4)
  &= \cS^{(\III,\x,\ell)}_{ik}(q_1,q_2,q_3,q_4) + \cS^{(\III,\x,\ell)}_{ik}(q_3,q_4,q_1,q_2) \;.
\end{align}
The right-hand side of Eq.~\eqref{WllLL} includes two types of mixed QCD$\times$QED contributions: one controlled by the function $\cS_{ikm}^{(\ell)}(q_1,q_2,q_3,q_4)$ which connects 3 hard partons and features charge asymmetry, the other controlled by the functions
$\cS_{ik}^{(\III,\cdots)}(q_1,q_2,q_3,q_4)$ which connects 2 hard partons and is charge symmetric.

\subsection{Soft emission of one quark pair and one lepton pair\label{s:qqll}}

Here we consider the case of radiation of one soft quark-antiquark pair $\qqbar$
and a soft lepton-antilepton pair $\LLbar$. Since leptons couple only to photons (in QED), the two-lepton soft current~\eqref{J1ll} is essentially a c-number and thus it commutes with the quark current~\eqref{J1ff} with $f\to\quark$, which we rewrite as
\begin{align}\label{J1qq}
  &\bj_{\qqbar}(q_1,q_2) = \bj(q_1,q_2) + \bj^{(\QED)}_{\qqbar}(q_1,q_2) \\
  &\bj^{(\QED)}_{\qqbar}(q_1,q_2) = - \left( \gq \,\mu^\ep \right)^2  \,e_\quark
  \;{\bf \Delta}_\quark \, \sum_{i} e_i \;\frac{p_i \cdot j(1,2)}{p_i \cdot q_{12}} \;.
\end{align}
The corresponding soft current can be written as
\begin{equation}\label{Jqqll}
  \bj_{\qqbar\LLbar}(q_1,q_2,q_3,q_4) =
  \bj_{\qqbar}(q_1,q_2)\bj^{\QED}_{\LLbar}(q_3,q_4)
  + \Ga_{\qqbar\LLbar}(q_1,q_2,q_3,q_4) \;.
\end{equation}

Since leptons interact only by photon exchange, we have two classes of diagrams contributing to the irreducible correlation: 1) diagrams A, B, C and D of fig.~\ref{f:qqQQ} with $\QQbar\to\LLbar$ and with two photons replacing the two gluons; 2) diagrams A and B with $\QQbar\to\LLbar$, a gluon attached to the hard leg and one photon connecting the quark line with the lepton line:
\begin{subequations}\label{Gqqll}
\begin{align}
  &\Ga_{\qqbar\LLbar}(q_1,q_2,q_3,q_4)
  = \Ga^{(\QED)}_{\qqbar\LLbar}(q_1,q_2,q_3,q_4) + \Ga^{(\MIX)}_{\qqbar\LLbar}(q_1,q_2,q_3,q_4) \\
  &\Ga^{(\QED)}_{\qqbar\LLbar}(q_1,q_2,q_3,q_4)
  \eqcs \left(\gq\,\mu^\ep\right)^4 \,e_\quark {\bf\Delta}_\quark \, e_L
  \sum_i e_i [e_\quark \gamma_i^{(\x)}(q_1,q_2,q_3,q_4) + e_L \gamma_i^{(\y)}(q_1,q_2,q_3,q_4)] \\
  &\Ga^{(\MIX)}_{\qqbar\LLbar}(q_1,q_2,q_3,q_4)
  \eqcs \left(\gq\,\g \,\mu^{2\ep}\right)^2 \,e_\quark\, \btq^c \,e_L
  \sum_i T^c_i \gamma_i^{(\x)}(q_1,q_2,q_3,q_4) \;.
\end{align}
\end{subequations}
Consequently, only the two ``abelian structures'' ($\sigma\in\{\x,\y\}$) contribute to the irreducible correlation $\Ga$ [cfr.~Eqs.~(\ref{G1234},\ref{R123})]; the kinematical coefficients
$\gamma_i^{(\sigma)}$ are the same of Eq.~\eqref{g12}.
Also in this case the dependence on the auxiliary gauge vector $n$ disappears
when the current $\bj_{\qqbar\LLbar}$ acts on colour singlet states.
Therefore the expressions given above are gauge-independent.

The squared current for soft $\qqbar\LLbar$ emission is computed by using the expressions in Eqs.~\eqref{Jqqll} and \eqref{Gqqll}. We write the result as follows
\begin{equation}\label{J2qqll}
  | \bj_{\qqbar\LLbar}(q_1,q_2,q_3,q_4) |^2
  = |\bj_{\qqbar}(q_1,q_2)|^2 \, |\bj^{\QED}_{\LLbar}(q_3,q_4)|^2
  + W_{\qqbar\LLbar}(q_1,q_2,q_3,q_4) \;,
\end{equation}
where $|\bj^{\QED}_{\LLbar}(q_3,q_4)|^2$ is provided by Eq.~\eqref{J2ll} with
$(1\to 3,2\to 4, e_\ell\to e_L)$ and
\begin{align}\label{J2qq}
  |\bj_{\qqbar}(q_1,q_2)|^2 &= |\bj(q_1,q_2)|^2 + |\bj^{\QED}_{\qqbar}(q_1,q_2)|^2 \nonumber \\
  &= \left(\g\,\mu^\ep \right)^4 T_R \sum_{i,k} \T_i\cdot\T_k \;{\cal I}_{ik}(q_1,q_2)
  + \left(\gq \,\mu^\ep \right)^4 N_c\, e_\quark^2 \sum_{i,k}  e_i\,e_k \;{\cal I}_{ik}(q_1,q_2)\;.
\end{align}
The irreducible correlation contribution $W_{\qqbar\LLbar}$ reads%
\footnote{Note the exchange $1\lra 3,2\lra 4$ in the momenta of some $\cS$ functions.}
\begin{align}
  W_{\qqbar\LLbar} &= -(\g\,\mu^{\ep})^4 (\gq\,\mu^{\ep})^4 \,T_R \Bigl\{
  2\,e_\quark e_L^2
    \sum_{i,k,m} \T_i\cdot\T_k\, e_m\, \cS^{(\ell)}_{ikm}(q_1,q_2,q_3,q_4) \nonumber \\
 & \hspace{9em} + e_\quark^2 e_L^2\sum_{i,k} \T_i\cdot\T_k\,
   \cS_{ik}^{(\III,\x,\ell)}(q_1,q_2,q_3,q_4) \Bigr\}\nonumber \\
 & - (\gq\,\mu^{\ep})^8 N_c \, e_\quark^2 e_L^2 \Bigl\{
   2 \sum_{i,k,m} e_i e_k e_m \bigl[e_\quark \cS^{(\ell)}_{ikm}(q_1,q_2,q_3,q_4)
   +e_L \cS^{(\ell)}_{ikm}(q_3,q_4,q_1,q_2) \bigr] \nonumber\\
 & \hspace{7em} + \sum_{i,k} e_i e_k
  \bigl[e_\quark^2 \cS_{ik}^{(\III,\x,\ell)}(q_1,q_2,q_3,q_4)
  +e_L^2 \cS_{ik}^{(\III,\x,\ell)}(q_3,q_4,q_1,q_2) \nonumber\\
 & \hspace{11em} +2 e_\quark e_L \cS_{ik}^{(\III,\y)}(q_1,q_2,q_3,q_4)
  \bigr] \Bigr\} \;. \label{Wqqll}
\end{align}

The pure leptonic irreducible correlation $W_{\llbar\LLbar}$ in Eq.~\eqref{WllLL} can be obtained from the
$\ord{\gq^8}$ term of Eq.~\eqref{Wqqll} by replacing $(N_c\to 1, e_\quark\to e_\ell)$.

\subsection{Soft emission of two quark-antiquark pairs\label{s:qqQQ}}

Here we consider the case of radiation of two soft quark-antiquark pairs $\qqbar$ in the presence of both gluons and photons. In this case the soft current reads
\begin{equation}\label{JqqQQ}
  \bj_{\qqbar\QQbar}(q_1,q_2,q_3,q_4) =
  \bj_{\qqbar}(q_1,q_2)\bj_{\QQbar}(q_3,q_4)
  + \Ga_{\qqbar\QQbar}(q_1,q_2,q_3,q_4) \;,
\end{equation}
where $\bj_{\qqbar}(q_1,q_2)$ is given in Eq.~\eqref{J1qq},
$\bj_{\QQbar}(q_3,q_4)$ is also given by Eq.~\eqref{J1qq} replacing $e_\quark\to e_\Quark$,
$q_1\to q_3$ and $q_2\to q_4$.
The irreducible current $\Ga_{\qqbar\QQbar}$ can be decomposed into four contributions:
\begin{equation}\label{GqqQQ}
  \Ga_{\qqbar\QQbar} = \Ga + \Ga^{(\MIX,H)}_{\qqbar\QQbar}
  + \Ga^{(\MIX,S)}_{\qqbar\QQbar} + \Ga^{(\QED)}_{\qqbar\QQbar} \;,
\end{equation}
where the first term $\Ga$ is the pure QCD irreducible current of Eq.~\eqref{G1234},
$\Ga^{(\MIX,H)}_{\qqbar\QQbar}$ represents the sum of diagrams A+B+C+D with a gluon attached to the hard leg and the photon between the soft quark lines,
$\Ga^{(\MIX,S)}_{\qqbar\QQbar}$ represents the sum of diagrams A+B+C+D with a gluon between the soft quark lines and the photon attached to the hard leg,
$\Ga^{(\QED)}_{\qqbar\QQbar}$ represents the sum of diagrams A+B+C+D with only photons instead of gluons. In formulae
\begin{align}\label{Gammas}
  \Ga^{(\MIX,H)}_{\qqbar\QQbar}(q_1,q_2,q_3,q_4) &\eqcs \left(\g\,\mu^\ep\right)^2
  \left(\gq\,\mu^\ep\right)^2\,
  \sum_i T^c_i \,\sum_{\sigma\in\{\x,\y\}} \cols_{(H,\sigma)}^c \gamma_i^{(\sigma)}(q_1,q_2,q_3,q_4) \\
  &\cols_{(H,\x)}^c = e_\quark e_\Quark t^c_{\alpha_1\alpha_2} \delta_{\alpha_3 \alpha_4} \;,\quad
   \cols_{(H,\y)}^c = e_\quark e_\Quark \delta_{\alpha_1 \alpha_2} t^c_{\alpha_3\alpha_4}
  \label{RH123} \\[2ex]
  \Ga^{(\MIX,S)}_{\qqbar\QQbar}(q_1,q_2,q_3,q_4) &\eqcs \left(\g\,\mu^\ep\right)^2
  \left(\gq\,\mu^\ep\right)^2\,
  \sum_i e_i \,\sum_{\sigma\in\{\x,\y\}} \cols_{(S,\sigma)} \gamma_i^{(\sigma)}(q_1,q_2,q_3,q_4) \\
  &\cols_{(S,\x)} = e_\quark t^b_{\alpha_1\alpha_2} t^b_{\alpha_3 \alpha_4} \;,\quad
   \cols_{(H,\y)} = e_\Quark t^b_{\alpha_1 \alpha_2} t^b_{\alpha_3\alpha_4}
  \label{RS123} \\[2ex]
  \Ga^{(\QED)}_{\qqbar\QQbar}(q_1,q_2,q_3,q_4) &\eqcs \left(\gq\,\mu^\ep\right)^4\,
  \sum_i e_i \,\sum_{\sigma\in\{\x,\y\}} \cols_{(Q,\sigma)} \gamma_i^{(\sigma)}(q_1,q_2,q_3,q_4) \\
  &\cols_{(Q,\x)} = e_\quark^2 e_\Quark \delta_{\alpha_1\alpha_2} \delta_{\alpha_3 \alpha_4} \;,\quad
   \cols_{(H,\y)} = e_\quark e_\Quark^2 \delta_{\alpha_1 \alpha_2} \delta_{\alpha_3\alpha_4}
  \label{RQ123} 
\end{align}
and the kinematic functions $\gamma_i^{(\sigma)}$ exactly equal to those in Eq.~\eqref{g12}.

The squared current for soft $\qqbar\QQbar$ emission is given by
\begin{equation}\label{J2qqQQ}
  | \bj_{\qqbar\QQbar}(q_1,q_2,q_3,q_4) |^2
  = \sy{ |\bj_{\qqbar}(q_1,q_2)|^2 }{ |\bj_{\QQbar}(q_3,q_4)|^2 }
  + W_{\qqbar\QQbar}(q_1,q_2,q_3,q_4) \;,
\end{equation}
where $|\bj_{\qqbar}(q_1,q_2)|^2$ is given by Eq.~\eqref{J2qq}, $|\bj_{\QQbar}(q_3,q_4)|^2$ is also obtained from Eq.~\eqref{J2qq} with the replacement $(1\to 3,2\to 4, e_\quark \to e_\Quark)$,
while the irreducible correlation contribution $W_{\qqbar\QQbar}$ is given by the pure QCD expression $W$ in Eq.~\eqref{Wqcd} [or in Eq.~\eqref{Wred}] plus an additional term $W'$ containing also photon interactions:
\begin{align}
  &W_{\qqbar\QQbar} = W + W' \label{WqqQQ} \\
  &W'(q_1,q_2,q_3,q_4) \eqcs \nonumber \\
  & -(\g\,\mu^{\ep})^6 (\gq\,\mu^{\ep})^2  \,2\,T_R^2 \Bigl\{
   e_\quark e_\Quark \sum_{i,k} \T_i\cdot\T_k\,
   \bigl[ \cS_{ik}^{(\III,\x,\ell)}(q_1,q_2,q_3,q_4)
   + \left(\substack{1\lra 3\\2\lra 4}\right) \bigr] \nonumber \\
  & \hspace{3em} + 
    \sum_{i,k,m} e_i\,\T_k\cdot\T_m\, \bigl[e_\quark \cS^{(\ell)}_{ikm}(q_1,q_2,q_3,q_4)
    + \left(\substack{1\lra 3\\2\lra 4\\e_\quark \lra e_\Quark}\right) +(i\lra k)\bigr]
   \Bigr\}\nonumber \\
  & -(\g\,\mu^{\ep})^4 (\gq\,\mu^{\ep})^4 \,T_R\,N_c \Bigl\{
   2 e_\quark e_\Quark \sum_{i,k,m} \T_i\cdot\T_k\, e_m \bigl[e_\Quark \cS^{(\ell)}_{ikm}(q_1,q_2,q_3,q_4)
    + \left(\substack{1\lra 3\\2\lra 4\\e_\quark \lra e_\Quark}\right) \bigr] \nonumber \\
  &\hspace{3em} + e_\quark^2 e_\Quark^2 \sum_{i,k} \T_i\cdot\T_k\,
   \bigl[\cS_{ik}^{(\III,\x,\ell)}(q_1,q_2,q_3,q_4)
   + \left(\substack{1\lra 3\\2\lra 4}\right) \bigr] \nonumber \\
  & \hspace{3em} + C_F \sum_{i,k} e_i e_k\,\Bigl[ \bigl[ e_\quark^2
   \cS_{ik}^{(\III,\x,\ell)}(q_1,q_2,q_3,q_4)
   + \left(\substack{1\lra 3\\2\lra 4\\e_\quark\lra e_\Quark}\right) \bigr]
  +2 e_\quark e_\Quark \cS_{ik}^{(\III,\y)}(q_1,q_2,q_3,q_4) \Bigr] 
  \Bigr\}\nonumber \\ 
  & -(\gq\,\mu^{\ep})^8 \,N_c^2 \, e_\quark^2 e_\Quark^2  \Bigl\{
   2\sum_{i,k,m} e_i e_k e_m \bigl[e_\quark \cS^{(\ell)}_{ikm}(q_1,q_2,q_3,q_4)
   + \left(\substack{1\lra 3\\2\lra 4\\e_\quark \lra e_\Quark}\right) \bigr] \nonumber\\
 & \hspace{3em} + \sum_{i,k} e_i e_k
  \Bigl[\bigl[e_\quark^2 \cS_{ik}^{(\III,\x,\ell)}(q_1,q_2,q_3,q_4)
  + \left(\substack{1\lra 3\\2\lra 4\\e_\quark\lra e_\Quark}\right) \bigr]
 +2 e_\quark e_\Quark \cS_{ik}^{(\III,\y)}(q_1,q_2,q_3,q_4)
  \Bigr] \Bigr\} \;. \nonumber
\end{align}

The quark-lepton irreducible correlation $W_{\qqbar\LLbar}$ in Eq.~\eqref{Wqqll} can be obtained
from the $\ord{\g^4\,\gq^4}$ and $\ord{\gq^8}$ terms of Eq.~\eqref{WqqQQ} by
removing in the $\ord{\g^4\,\gq^4}$ term the $C_F$ part and removing also the $(\quark\lra\Quark)$ ``symmetrized'' contributions, reducing by one each power of $N_c$, i.e., $N_c^k \to N_c^{k-1}$ and replacing everywhere $e_\Quark \to e_L$.

\section{Summary\label{s:sum}}

We have considered the radiation of two distinguishable soft quark-antiquark pairs ($\qqbar\QQbar$) in QCD hard scattering. In this soft limit, the scattering amplitude exhibits a singular behaviour that is factorised and controlled by a multiparton soft current, which possesses a process-independent structure in colour space.

We have evaluated the soft $\qqbar\QQbar$ current at tree level for a generic scattering amplitude with an arbitrary number of external hard partons. The soft current acts in colour space and is expressed in terms of the colour charges and momenta of the external hard partons. We have decomposed the current into a contribution arising from the symmetrised product of independent single-pair emissions and an irreducible correlation term. This irreducible component includes strictly non-abelian contributions as well as terms with an abelian character, reflecting the structure observed in the soft gluon-quark-antiquark ($g\qqbar$) case but extended to the four-fermion topology.

We have computed the tree-level squared current $|\bj(q_1, q_2, q_3, q_4)|^2$ and the ensuing colour correlations for squared amplitudes of generic multiparton hard-scattering processes. The squared current leads to two main types of colour interactions between the hard partons: the customary colour dipole interactions and colour tripole interactions proportional to the fully-symmetric tensor $d^{abc}$. This structure offers a compelling comparison with the case of triple soft-gluon radiation. In the three-gluon case, the irreducible correlations are governed by colour quadrupole operators, which possess a maximally non-abelian character involving the contraction of structure constants, and preserve the charge-conjugation invariance of the squared amplitude. In contrast, for the emission of two soft quark-antiquark pairs, the dominant non-dipolar correlations are of the tripole type. These tripole interactions are odd under charge conjugation and are responsible for the emergence of charge asymmetry effects between the soft fermions and antifermions. This feature is shared with the case of soft $g\qqbar$ emission, confirming that charge asymmetry is a distinctive signature of soft-fermion radiation that is absent in pure soft-gluon emission at tree level.

We have carried out numerical comparisons between our results and those of 
Ref.~\cite{Chen:2024hvp}. We find agreement for the terms with the symmetric product of uncorrelated $\qqbar$ squared currents and also for the tripole contribution, but we find disagreement for the dipole-correlated terms.

We have explicitly applied our general results to the specific cases of processes with two and three hard partons. In these configurations, the colour structure can be partially simplified. For processes with two hard partons, the factorisation reduces to a c-number form. For processes with three hard gluons, we discussed how the squared current acts as a matrix in the colour space of the hard partons, inducing transitions between colour-symmetric and colour-antisymmetric states via the tripole operators.

Finally, we have generalised our QCD analysis to include QED and mixed QCD$\times$QED interactions. We have presented the corresponding results for the emission of two soft lepton-antilepton pairs ($\ell\bar{\ell}L\bar{L}$), for the mixed emission of a quark-antiquark pair and a lepton-antilepton pair ($\qqbar\LLbar$) and also for $\qqbar\QQbar$ pairs. We observed that, while pure QED emission retains an abelian character with no charge asymmetry in the squared current, the mixed QCD$\times$QED contributions can induce non-trivial correlations dependent on the specific parton content.

The results presented in this paper are of value regarding soft-parton emission at $\mathcal{O}(\alpha_S^4)$ (and mixed orders) at tree level, providing necessary ingredients for high-precision calculations at N$^3$LO and N$^4$LO, and for the resummation of large logarithmic corrections in collider observables.

\section*{Acknowledgments}

\includegraphics[width=2.2em,angle=90]{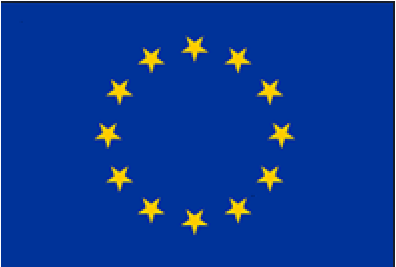}~
\begin{minipage}[b]{0.9\linewidth}
  This project has received funding from the European Union’s Horizon
  2020 research and innovation programme under grant agreement No
  824093. 
\end{minipage}
LC is supported by the Generalitat Valenciana (Spain) through the plan GenT program (CIDEGENT/2020/011) and his work is supported by the Spanish Government (Agencia Estatal de Investigación) and ERDF funds from European Commission (Grant no. PID2020-114473GB-I00 funded by MCIN/AEI/10.13039/501100011033).


\end{document}